\documentclass[aps,prb,twocolumn,superscriptaddress]{revtex4-1}
\usepackage{mathrsfs}
\usepackage{graphicx}
\usepackage{amsmath,amsfonts,amssymb}
\usepackage{color}
\usepackage{ulem}
\usepackage{subfigure}
\usepackage{physics}
\usepackage{dsfont}
\usepackage{hyperref}
\usepackage{url}
\usepackage{leftidx}
\usepackage{textcomp}
\usepackage{gensymb}

\definecolor{emerald}{rgb}{0.31, 0.78, 0.47}
\definecolor{blue(ncs)}{rgb}{0.0, 0.53, 0.74}

\hypersetup{
     colorlinks=true,
     linkcolor=blue,
     filecolor=blue,
     citecolor=emerald,      
     urlcolor=blue(ncs),
}

\usepackage{chngcntr}

\DeclareMathAlphabet{\pazocal}{OMS}{zplm}{m}{n}

\newcommand{\bo}[1]{\boldsymbol{#1}}

\newcommand{\D}{\mathrm{d}}
\newcommand{\f}[2]{\frac{#1}{#2}}

\begin{document}
\normalem

\title{A microscopic Ginzburg--Landau theory and singlet ordering in Sr$_2$RuO$_4$}
  
\author{Glenn Wagner}
  \affiliation{Rudolf Peierls Center for Theoretical Physics, Oxford OX1 3PU, United Kingdom}
  \affiliation{Kavli Institute for Theoretical Physics, University of California Santa Barbara, CA 93106, USA}
  
\author{Henrik S. R{\o}ising}
 \email{henrik.roising@su.se}
  \affiliation{Rudolf Peierls Center for Theoretical Physics, Oxford OX1 3PU, United Kingdom}
  \affiliation{Nordita, KTH Royal Institute of Technology and Stockholm University, Hannes Alfv\'{e}ns v\"{a}g 12, SE-106 91 Stockholm, Sweden}
  
\author{Felix Flicker}
  \affiliation{Rudolf Peierls Center for Theoretical Physics, Oxford OX1 3PU, United Kingdom}
  \affiliation{School of Physics and Astronomy, Cardiff University, Cardiff CF24 3AA, United Kingdom}
  \affiliation{School of Mathematics, University of Bristol, Bristol BS8 1TW, United Kingdom}
  
\author{Steven H. Simon}
  \affiliation{Rudolf Peierls Center for Theoretical Physics, Oxford OX1 3PU, United Kingdom}
  
\date{\today}

\begin{abstract}
The long-standing quest to determine the superconducting order of Sr$_2$RuO$_4$ (SRO) has received renewed attention after recent nuclear magnetic resonance (NMR) Knight shift experiments have cast doubt on
the possibility of spin-triplet pairing in the superconducting state. As a putative solution, encompassing a body of experiments conducted over the years, a $(d+ig)$-wave order parameter caused by an accidental near-degeneracy has been suggested [S.~A.~Kivelson et al., npj Quantum Materials $\bo{5}$, 43 (2020)]. Here we develop a general Ginzburg--Landau theory for multiband superconductors. We apply the theory to SRO and predict the relative size of the order parameter components. The heat capacity jump expected at the onset of the second order parameter component is found to be above the current threshold deduced by the experimental absence of a second jump. Our results tightly restrict theories of $d+ig$ order, and other candidates caused by a near-degeneracy, in SRO. We discuss possible solutions to the problem.
\end{abstract}

\maketitle

%
%%
%%%
\section{Introduction}
\label{sec:Intro}
%%%
%%
%
$26$ years ago the layered perovskite Sr$_2$RuO$_4$ (SRO) was found to harbour unconventional superconductivity below the modest critical temperature $T_c \approx 1.5$~K~\cite{MaenoEA94}. Its superconducting order was widely believed to be chiral $p$-wave~\cite{MackenzieMaeno03}. This belief was primarily rooted in the absence of a drop in the nuclear magnetic resonance (NMR) Knight shift~\cite{IshidaEA98}, and the indications of time-reversal symmetry-breaking (TRSB) found in muon spin relaxation~\cite{Luke98} ($\mu$SR) and Kerr rotation~\cite{XiaEA06} experiments. Chiral $p$-wave superconductors open the possibility of hosting Majorana zero modes which have intriguing applications to topological quantum computation~\cite{NayakEA08}.

Over the years, the number of experimental results not conforming with the chiral $p$-wave hypothesis have accumulated~\cite{MackenzieEA17}. Among observations difficult to explain within the chiral $p$-wave paradigm are indications of gap nodes inferred from heat capacity~\cite{NishiZakiEA00, GrafBalatsky00}, heat conductivity~\cite{HassingerEA17} and scanning tunneling microscopy measurements (STM)~\cite{MadhavenEA19}, and the absence of a $T_c$-cusp under uniaxial strain~\cite{HicksEA14, SteppkeEA17}. A peak in the accumulated evidence was reached when the NMR Knight shift experiment was repeated~\cite{PustogowEA19, IshidaEA19}, now finding a substantial reduction in the spin susceptibility at low temperature. This has launched a renewed focus on the compound, both experimentally~\cite{LiEA19, PetchEA20, GrinekoEA20, GhoshEA20, ChronisterEA20, XinxinEA20} and theoretically~\cite{GingrasEA19, RomerEA19, RoisingEA19, LindquistKee19, SuhEA19, ZhiqiangEA20, KivelsonEA20, Mazumdar20, RomerEA20, Willa20, LeggettEA20}. 

The new NMR experiments~\cite{PustogowEA19, IshidaEA19} appear reconcilable with a number of even-parity pseudospin singlet order parameters and possibly the helical $p$-wave pseudospin triplet order parameters. However, the options are being narrowed down as thermodynamic shear elastic measurements~\cite{GhoshEA20} and $\mu$SR~\cite{GrinekoEA20} suggest that the superconducting order is likely two-component, at least at temperatures well below $T_c$. A very recent NMR experiment at low magnetic fields casts further doubt on odd-parity order~\cite{ChronisterEA20} (restricting any odd-parity component to be $\lesssim 10$\% of the primary component), leaving pseudospin-singlet pairing as the most likely scenario. Recently, two putative solutions to the long-standing puzzles have been proposed. 

A chiral $d$-wave order parameter (irreducible representation (irrep.) $E_g$ of $D_{4h}$ in the group theory nomenclature) could explain TRSB and the observed jump in the shear elastic modulus $c_{66}$~\cite{GhoshEA20}. Indeed the behaviour of $T_c$ and $T_\textrm{TRSB}$ under both hydrostatic pressure and La substitution \cite{splitting_Tc_TTRSB} is similar, suggesting a symmetry protected degeneracy  such as this one. It was shown that a chiral $d$-wave can be stabilized by including certain $k_z$-dependent spin-orbit coupling (SOC) terms at sufficiently large Hund's coupling~\cite{SuhEA19}. However, the prevailing belief has been that the material is effectively two-dimensional (2D)~\cite{BergmannEA00, MackenzieEA17}, a belief which has recently been examined and to some extent confirmed~\cite{DamascelliEA14, GingrasEA19,  RoisingEA19}. Furthermore, the horizontal line node that the $d_{xz} +id_{yz}$ order possesses, would likely conflict with the experimental evidence of vertical line nodes~\cite{HassingerEA17, MadhavenEA19}. 

Another possibility, solving the latter issue, is an accidental (near-)degeneracy between a $d_{x^2-y^2}$ and $g_{xy(x^2-y^2)}$-wave order parameter~\cite{KivelsonEA20}. This scenario has the potential of explaining both features of the temperature vs.~strain phase diagram, indications of TRSB, vertical line nodes, and the shear elastic modulus jump. However, although various theories find $d$-wave order as the leading instability~\cite{GingrasEA19, RomerEA19, RoisingEA19, ZhiqiangEA20, RomerEA20}, an exotic $g$-wave order becoming competitive currently lacks support from calculations using the relevant band structure. Moreover, an accidental near-degeneracy would imply the presence of a secondary, possibly small, heat capacity jump at a temperature $T_{\text{TRSB}} < T_c$. Despite intensive search for a second jump in high-precision measurements~\cite{LiEA19}, such an observation remains elusive. On the other hand, such a secondary heat capacity jump \textit{has} been observed for the multicomponent superconductor UPt$_3$, which is believed to have  chiral $f$-wave order~\cite{FisherEA89, JoyntEA02, KallinEA16}.

Here we address the feasibility of a $d+ig$ order parameter in SRO by taking on a microscopic perspective to discuss the heat capacity anomaly. We first develop the framework for a general multiband, multi-component Ginzburg--Landau (GL) theory where the expansion coefficients depend on the band structure. Our theory reduces to that of Gor'kov~\cite{Gorkov59} for quadratic bands and a single-component $s$-wave order parameter. Using band and gap structures applicable to SRO we find, by numerical minimization of the free energy, that the $g$-wave component prefers to have a magnitude of about $71\%$ of the $d$-wave component at low temperature. We calculate the expected secondary heat capacity jump and evaluate it numerically as a function of the order parameter component sizes. The results predict a second jump larger than what is seen experimentally, meaning fine-tuning would be required in any possible $d+ig$ scenario. The same conclusion is reached for other near-degeneracy options.

Finally, variations of the general theory developed here could also prove to have applications to exotic (chiral) superconductors~\cite{Ghosh2EA20} outside the scope of SRO, like FeAs-based systems~\cite{LeeEA09}, UTe$_2$~\cite{RanEA19, MadhavenEA20}, and URu$_2$Si$_2$~\cite{PalstraEA85, SchemmEA15}.

%
%%
%%%
\section{Theory: Multiband Ginzburg--Landau}
\label{sec:Derivation}
%%%
%%
%
In this section we develop a generic expansion of the free energy in the order parameter close to the critical temperature for a multiband superconductor. We initiate the approach for a general multi-component order parameter on the lattice. We vindicate the theory in the case of a single-component $s$-wave order parameter for quadratically dispersing bands, for which we reproduce well-established results~\cite{Gorkov59}. Then we consider the case of two nearly degenerate pseudospin singlet order parameter components.

\subsection{General formalism}
\label{sec:Formalism}
\begin{figure}[b!th]
	\centering
	\subfigure[]{\includegraphics[width=0.24\linewidth]{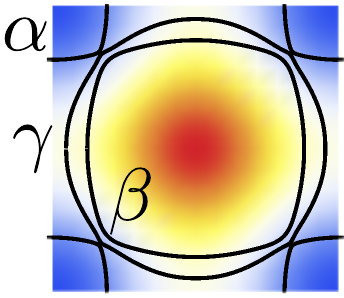}} \quad 
	\subfigure[]{\includegraphics[width=0.20\linewidth]{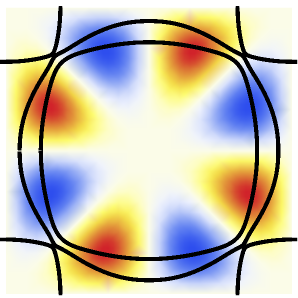}} \quad
		\subfigure[]{\includegraphics[width=0.20\linewidth]{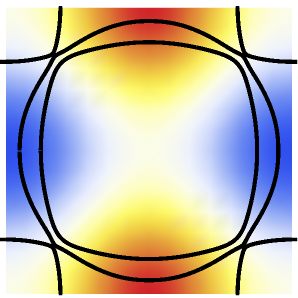}} \quad 
	\subfigure[]{\includegraphics[width=0.20\linewidth]{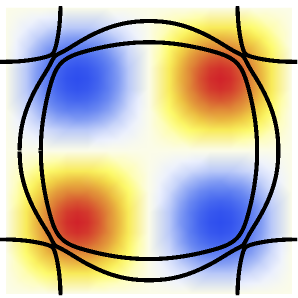}} 
\caption{Symmetries of the even-parity order parameters in channel (a) $A_{1g}$, (b) $A_{2g}$, (c) $B_{1g}$, (d) $B_{2g}$. The black lines display the Fermi surface applicable to the three-band case of Sr$_2$RuO$_4$, where the three bands are denoted by $\alpha$, $\beta$, and $\gamma$. The Fermi surface is obtained using tight-binding parameters listed in Appendix~\ref{sec:TBModel}, as extracted from density functional theory~\cite{SteppkeEA17, Rosner}.}
	\label{fig:OPExamples}
\end{figure}

We start with a single-particle tight-binding Hamiltonian in orbital/spin space. Due to the presence of spin-orbit coupling we transform to the band/pseudospin basis in which the Hamiltonian is diagonal,
\begin{equation}
    H_{\text{N}} = \sum_{\mu, \sigma, \bo{p}} \xi_{\mu}(\bo{p}) c_{\mu\sigma}^{\dagger}(\bo{p}) c_{\mu \sigma}(\bo{p}). 
    \label{eq:Hkin} 
\end{equation}
Above, $\xi_{\mu}(\bo{p})$ is the dispersion of band $\mu$ ($\mu = \alpha, \beta, \gamma$ in the case of SRO), and $\sigma = \Uparrow, \Downarrow$ denotes pseudospin, with $\bar{\sigma}$ being the opposite pseudospin of $\sigma$. The sum over $\bo{p}$ runs over the first Brillouin zone. $c_{\mu\sigma}^{\dagger}(\bo{p})$ creates an electron in band $\mu$ with pseudospin $\sigma$. See Appendix \ref{sec:TBModel} for further details of the non-interacting Hamiltonian. In this work we choose to focus on pseudospin singlet pairing. The pseudospin singlets that we find will have a spin-triplet component, which, however, is small~\footnote{As mentioned in the introduction, new NMR measurements~\cite{ChronisterEA20}, going down to magnetic fields of $B < 0.2 B_{c2}$ at $T = 25$mK, have constrained any spin-triplet component to be less than about $10\%$ of the spin-singlet component. The size of the spin-triplet component in the pseudospin singlets is dictated by the strength of the spin-orbit coupling ($\lambda$ in Eq.~\eqref{eq:KineticHamiltonian}) in the transformation $c_{as}(\bo{p}) = \sum_{\mu, \sigma} u_{as}^{\mu \sigma}(\bo{p}) c_{\mu \sigma}(\bo{p})$, where $u_{as}^{\mu \sigma}(\bo{p})$ is an eigenvector component of $h_s(\bo{p})$ in Eq.~\eqref{eq:KineticHamiltonian}. }.
\begin{table}[t!bh]
\caption{One-dimensional, even-parity (pseudospin singlet) irreducible representations of the tetragonal point group $D_{4h}$~\cite{SigristUeda91}. Lattice harmonics of order parameters are listed in the Balian--Werthamer basis~\cite{BalianWerthamer63}, $\Delta_{\sigma \sigma'} = [ i d_0(\theta) \sigma_y]_{\sigma \sigma'}$, where $\theta$ is the polar angle (2D)~\cite{SimkovicEA16}.}
\begin{center}
\begin{tabular}{p{1.0cm} p{1.5cm} p{4.0cm} }
\toprule
 Irrep. & Name & Lattice harmonics of $d_0(\theta)$ \\ \hline
 $A_{1g}$  & $s'$ & $\sum_{n = 1}^{\infty} a_{n} \cos(4n\theta)$ \\
 $A_{2g}$  & $g_{xy(x^2-y^2)}$ & $\sum_{n = 0}^{\infty} b_{n} \sin([4n+4]\theta)$ \\
 $B_{1g}$  & $d_{x^2-y^2}$ & $\sum_{n = 0}^{\infty} c_{n} \cos([4n+2]\theta)$ \\
 $B_{2g}$  & $d_{xy}$ & $ \sum_{n = 0}^{\infty} d_{n} \sin([4n+2]\theta)$ \\ \hline \hline
\end{tabular}
\end{center}
\label{tab:Representations}
\end{table}
We shall consider the pseudospin-singlet Cooper pairing terms $H_{\text{SC}}$ as perturbations to the normal-state Hamiltonian $H_{\text{N}}$ close to the critical temperature, where
\begin{equation}
\begin{aligned}
H_{\text{SC}} &= \sum_{\mu, \sigma,a }  {\sum_{\bo{p}}} \big[ \Delta_{a\mu}(\bo{p})  c_{\mu \sigma}^{\dagger}\left(\bo{p}\right) c_{\mu \bar{\sigma}}^{\dagger} \left(-\bo{p} \right) + \text{h.c.} \big]. 
\end{aligned}
\label{eq:HSC}
\end{equation}
Here $\Delta_{a\mu}(\bo{p})$ is the pseudospin-singlet order parameter of band $\mu$ corresponding to irrep.~$a$.  The sum over $\bo{p}$ runs over the Fermi surface sheet $\lvert \xi_{\mu} (\bo{p}) \rvert < \omega_c \sim k_BT$, where $\omega_c \ll W$ is an electronic cutoff small compared to the bandwidth $W$. Considering only intra-band terms is justified if the superconducting gap is small compared to the energy separation of the bands at the Fermi level, which indeed is satisfied in SRO where these energy scales are on the order of $0.5$~meV~\cite{MadhavenEA19} and $100$~meV~\cite{DamascelliEA14}, respectively. We shall focus on the tetragonal point group $D_{4h}$, for which the relevant one-dimensional irreps are listed in Table~\ref{tab:Representations} and visualised in Fig.~\ref{fig:OPExamples}.

Proceeding with the Ginzburg--Landau (GL) approach we expand the free energy density in the (multi-component) order parameter close to the critical temperature~\cite{Gorkov59} (see also Refs.~\onlinecite{SilaevEA12, Stanev2010,Maiti2013, BruusFlensberg2004, RupertEA16, LeeEA09, ZhiqiangEA20}). We assume that the  
critical temperature of the order parameter in irrep.~$a$ is $T_{ca}$. As the superconducting phase is entered, the corrections to the normal state free energy,  $\Delta F = F_{\text{SC}} - F_{\text{N}}$, are caused by the superconducting terms of Eq.~\eqref{eq:HSC}. The corrections can be evaluated using the Gibbs average of the $S$-matrix~\cite{AbrikosovGorkovDzyal58, Gorkov59, Sadovskii19},
\begin{figure}[t!bh]
	\centering
	\includegraphics[width=0.45\linewidth]{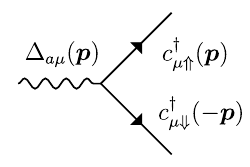}
\caption{The pseudospin-singlet Cooper pair operator shown diagrammatically as a two-fermion composite operator.}
	\label{fig:Cooper}
\end{figure}
\begin{figure*}[t!bh]
	\centering
	\subfigure[]{\includegraphics[width=0.55\linewidth]{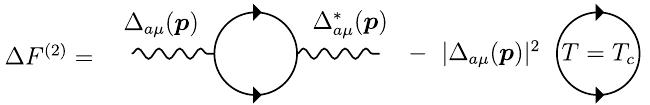}} \quad 
	\subfigure[]{\includegraphics[width=0.38\linewidth]{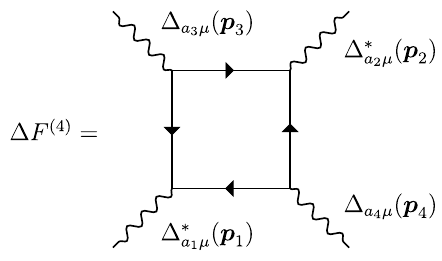}} 
\caption{The (a) second order and (b) fourth order diagrams contributing to the free energy of Eq.~\eqref{eq:LoopExpansion}. The algebraic expressions corresponding to panel (a) and (b) are given in Eq.~\eqref{eq:SecondOrderF} and \eqref{eq:FourthOrderF}, respectively. The single-component, quadratic band case resulting in Eqs.~\eqref{eq:MainSecondOrderF2} and \eqref{eq:MainFourthOrderF2} corresponds to fixing $a_1 = a_2 = a_3 = a_4 = A_{1g}$ here.}
	\label{fig:ContributionsF}
\end{figure*}
\begin{align}
\Delta F &= -T \ln\langle S \rangle, \label{eq:FreeEnergyCorrection} \\
S  &= \pazocal{T}_{\tau} \exp\big( -\int_{0}^{\beta} \D\tau~H_{\text{SC}} (\tau)  \big), \label{eq:Smatrix} 
\end{align}
where $\beta = 1/T$ (with $k_B = 1$), $\tau = -i \beta$ is imaginary time, and $\pazocal{T}_{\tau}$ is the time-ordering operator. The loop expansion of $\Delta F$, involving only connected diagrams, is given by
\begin{equation}
\begin{aligned}
\Delta F &= -T\big( \langle S\rangle_c - 1 \big) \\
&\approx \f{T}{2!} \int_{0}^{\beta}\D\tau_1 \int_{0}^{\beta}\D\tau_2~\big\langle \pazocal{T}_{\tau}\left[ H_{\text{SC}}(\tau_1) H_{\text{SC}}(\tau_2) \right]  \big\rangle_c \\
& + \f{T}{4!} \int_{0}^{\beta}\D\tau_1 \cdots \int_{0}^{\beta}\D\tau_4 ~\big\langle \pazocal{T}_{\tau}\left[ H_{\text{SC}}(\tau_1) \cdots H_{\text{SC}}(\tau_4) \right]  \big\rangle_c.
\end{aligned}
\label{eq:LoopExpansion}
\end{equation}
Pictorially this consists of closed connected diagrams with only external $\Delta$ legs produced by combinations of the Feynman diagram of Fig~\ref{fig:Cooper}.

To calculate the first and second terms of Eq.~\eqref{eq:LoopExpansion} for a weakly coupled superconductor (which is valid near the critical temperature), bare Green's functions are introduced as~\cite{BruusFlensberg2004}
\begin{equation}
G_{\mu}(\bo{p}, \tau_1-\tau_2) = - \big\langle \pazocal{T}_{\tau} c_{\mu \Uparrow}(\bo{p}, \tau_1) c_{\mu \Uparrow}^{\dagger}(\bo{p}, \tau_2) \big\rangle.
\label{eq:GreensFunction}
\end{equation}
This can be expressed in the Matsubara representation: $G_{\mu}(\bo{p}, \omega_n) = 1/(i\omega_n-\xi_{\mu}(\bo{p}))$, with fermionic Matsubara frequencies $\omega_n = \f{\pi}{\beta}(2n+1)$ for integer $n$. We evaluate the second and fourth order contributions of Eq.~\eqref{eq:LoopExpansion}, with corresponding diagrams shown in Fig.~\ref{fig:ContributionsF} (a) and (b), and find the free energy, $\Delta F =\Delta F^{(2)}+\Delta F^{(4)}$,
\begin{widetext}
\begin{align}
\label{eq:GLTheoryOne}\Delta F&= \sum_{\mu} \Big( {\sum_{a,\bo{p}}} \alpha_{a\mu}(\bo{p},T) \lvert \Delta_{a\mu}(\bo{p})\rvert^2 + {\sum_{a_i,\bo{p}_i}} \beta_{\{a_i\}\mu}(\{\bo{p}_i\},T) \Delta^{*}_{a_1\mu}(\bo{p}_1) \Delta^{*}_{a_2\mu}(\bo{p}_2) \Delta_{a_3\mu}(\bo{p}_3) \Delta_{a_4\mu}(\bo{p}_4) \Big),
\\
\alpha_{a\mu}(\bo{p},T) &= - T   \sum_{n} G_{\mu}(\bo{p}, \omega_n) G_{\mu}(-\bo{p}, -\omega_n)  + T_{ca}  \sum_{n} G_{\mu}(\bo{p}, \omega_n) G_{\mu}(-\bo{p}, -\omega_n) \rvert_{T = T_{ca}}, \label{eq:SecondOrderF} \\
\beta_{\{a_i\}\mu}(\{\bo{p}_i\},T) &= \f{T}{2} f_{a_1a_2a_3a_4} \delta_{\bo{p}_1,\bo{p}_3}\delta_{\bo{p}_1,\bo{p}_4} \delta_{\bo{p}_2,\bo{p}_3}\delta_{\bo{p}_2,\bo{p}_4} \sum_{n} G_{\mu}(\bo{p}_1, \omega_n)G_{\mu}(\bo{p}_2, \omega_n) G_{\mu}(-\bo{p}_3, -\omega_n)G_{\mu}(-\bo{p}_4, -\omega_n), \label{eq:FourthOrderF} \\
f_{a_1a_2a_3a_4} &\equiv \delta_{a_1a_3}\delta_{a_2a_4}+\delta_{a_1a_4}\delta_{a_2a_3}+\delta_{a_1a_2}\delta_{a_3a_4}-2\delta_{a_1a_2}\delta_{a_2a_3}\delta_{a_3a_4}. \label{eq:KroneckerCombination}
\end{align}
\end{widetext}
In $\alpha_{a\mu}(\bo{p},T) $ we subtracted off the contribution evaluated at $T_{ca}$ to ensure that $\Delta F$ has a well-defined minimum for $T < T_{ca}$.

\subsection{Specific limit}
\label{sec:Limits}
In this section we consider a specific limit of the expression for the free energy derived above, and we verify previously-established results in this limit. The details are listed explicitly in Appendix~\ref{sec:MmtmIndepGL}, we summarize the results here. 

To verify the theory we consider the simplifying case of (i) assuming a single-component $s$-wave order parameter, and (ii) quadratic bands in two dimensions. The assumption (i) amounts to setting $\Delta_{a \mu}(\bo{p}) = \Delta_{A_{1g} \mu} \equiv \Delta_{\mu}$. This allows us to pull the order parameters in Eq.~\eqref{eq:GLTheoryOne} outside the $\bo{p}$ sums and perform the Matsubara sums analytically. The resulting functions are sharply peaked around the Fermi surface, and the $\bo{p}$ sums can be converted to integrals which can be evaluated in closed form for quadratic bands. The final result for quadratically dispersing bands, $\xi_{\mu}(\bo{p}) = \bo{p}^2/(2m_{\mu})$, is
\begin{align}
    \Delta F
    &= \sum_{\mu} \Big( \tilde\alpha_{\mu}(T,T_c) \lvert \Delta_{\mu}\rvert^2  +  \tilde\beta_{\mu}(T) \lvert \Delta_{\mu}\rvert^4 \Big), \label{eq:DeltaF} \\
\tilde\alpha_{\mu}(T,T_c) &= \f{\rho_{\mu}}{2}\left(\f{T}{T_c} - 1 \right), \label{eq:MainSecondOrderF2} \\
\tilde\beta_{\mu}(T) &= \f{\rho_{\mu}}{T^2} \f{7\zeta(3)}{32\pi^2},\label{eq:MainFourthOrderF2}
\end{align}
with $\rho_{\mu} = V m_{\mu}/(2\pi)$ being the density of states, $\zeta$ is the Riemann zeta function, and where we assumed that $T/T_c-1 \ll 1$. This is equivalent to the result of Gor'kov~\cite{Gorkov59}.

In the more general case we assume that $\Delta_{a\mu}(\bo{p})=\Delta_{a}^0\Delta_{a\mu}(\bo{p})$. Here, $\Delta_{a\mu}(\bo{p})$ are normalized order parameters belonging to irrep.~$a$ of the crystal point group, and $\Delta_{a}^0$ are the amplitudes of a given irrep, which are the variational parameters over which we want to minimize our free energy. We note that these variational parameters do not depend on the band label $\mu$ since the relative amplitude of the gaps of a given irrep on the different bands is assumed fixed in $\Delta_{a \mu}(\bo{p})$. The free energy  becomes
\begin{align}
\Delta F&= {\sum_{a}} \tilde\alpha_{a}(T,T_{ca}) \lvert \Delta_{a}^0\rvert^2  + {\sum_{a_i}} \tilde\beta_{\{a_i\}}(T) \Delta_{a_1}^{0*} \Delta_{a_2}^{0*} \Delta_{a_3}^{0} \Delta_{a_4}^{0},
\end{align}
where the expressions for the GL coefficients $\tilde \alpha_{a}$ and $\tilde \beta_{\{a_i\}}$ are found in Appendix~\ref{sec:MmtmIndepGL}.
%
%
%%
%%%
\section{Application to SRO}
\label{sec:ApplicationSRO}
%%%
%%
%
In this section we apply the theory developed in Sec.~\ref{sec:Derivation} to the multiband case of SRO. In Sec.~\ref{sec:MinimizationSRO} we calculate the temperature-dependent order parameter weights. This is contrasted with the calculation of Sec.~\ref{sec:Anomaly}, where we estimate the heat capacity jump expected at the onset of a second order parameter component as a function of the two component sizes.

\subsection{The relative order parameter weight}
\label{sec:MinimizationSRO}
To minimize the free energy of Eq.~\eqref{eq:GLTheoryOne}, we employ the gap ans{\"a}tze listed in Tab.~\ref{tab:Representations}. Specifically, we fit the ans{\"a}tze to the order parameters resulting from the microscopic weak-coupling RG calculation of Ref.~\onlinecite{RoisingEA19}, thereby including order parameter anisotropies expected for SRO (see Appendix~\ref{sec:AdvancedOP} for details). For the band structure we work with a (2D) three-band model (which includes spin-orbit coupling), based on density functional theory~\cite{SteppkeEA17, Rosner}. This model is presented in Appendix~\ref{sec:TBModel}. We feed in the band structure of the $\alpha$, $\beta$, and $\gamma$ bands and evaluate Eqs.~\eqref{eq:SecondOrderF} and \eqref{eq:FourthOrderF} numerically using Monte-Carlo integration with the two-parameter theory $\Delta F[\lbrace  \Delta_0 \Delta_{a\mu}(\bo{p}), i \Delta_0 X \Delta_{b \mu}(\bo{p}) \rbrace ] \equiv\Delta F[\Delta_0, X] $. The function $\Delta F[\Delta_0, X]$ is minimized over the two scalar arguments: the overall gap size $\Delta_0$ and the relative weight $X$ as a function of temperature. In the two-parameter theory the $d+ig$ hypothesis is addressed by specifying $a = B_{1g}$ and $b = A_{2g}$. In addition to Ref.~\onlinecite{RoisingEA19} several other RG calculations have been performed~\cite{ScaffidiEA14, DodaroEA18, ZhangEA18, RomerEA19, SuhEA19, WangEA19, RomerEA21}, finding slightly different competing order parameters. However, one should not expect the shape of $\Delta_{a\mu}(\bo{p})$ for a given $a$ and $\mu$ to be vastly different in the multiple different approaches.

The resulting form of $X(T)$ determined from minimization of $\Delta F$ is shown in Fig.~\ref{fig:RGWeights}, employing a realistic three-band dispersion described in Appendix~\ref{sec:TBModel}, with the corresponding Fermi surface shown in Fig.~\ref{fig:OPExamples}.
\begin{figure}[t!bh]
	\centering
	\includegraphics[width=0.98\linewidth]{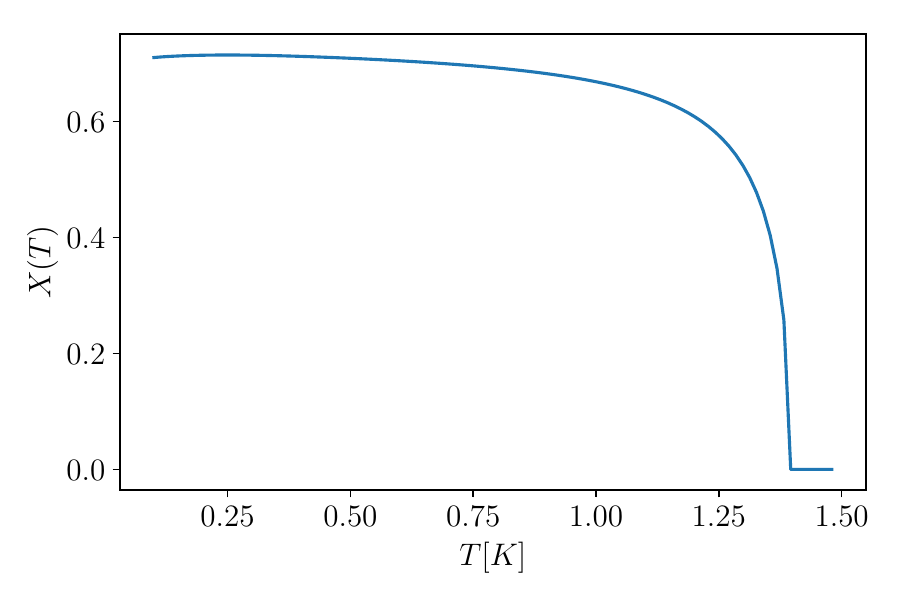}
\caption{The weight $X(T)$, with an order parameter of the form $\Delta(\bo{p}) = \Delta_0(T) [\Delta_{B_{1g} \mu}(\bo{p}) + i X(T) \Delta_{A_{2g} \mu}(\bo{p}) ]$ for $T_{c1}=1.48$~K and $T_{c2} = 1.44$~K. This result is obtained using the GL coefficients Eqs.~\eqref{eq:AppAlpha} and \eqref{eq:AppBeta}.}
\label{fig:RGWeights}
\end{figure}
The value of $X$ quickly tends to a value $>0.6$ as the temperature is lowered through $T_{\text{TRSB}} < T_{c2}$. 

\subsection{The heat capacity anomaly}
\label{sec:Anomaly}

In a recent $\mu$SR experiment~\cite{GrinekoEA20} two temperature scales were probed under uniaxial strain: $T_c$ and $T_{\text{TRSB}}$ as determined from the heat capacity jump and the abrupt change in the muon spin relaxation rate, respectively. The results indicate that (i) there is a sharp onset of TRSB at $T_{\text{TRSB}} \lesssim T_c$ (with $T_{\text{TRSB}}/T_c \approx 0.94$ when averaged over four samples), and (ii) that the two temperatures split increasingly under uniaxial strain.

However, measurements of the heat capacity resolved under uniaxial strain did not observe any secondary heat capacity jump, as would be expected with the onset of a second order parameter component~\cite{LiEA19, GrinekoEA20, BastianSigrist20}. This resulted in the experimental bound, deduced from the measurement resolution~\cite{GrinekoEA20, LiEA19}, that any secondary jump would have to be less than about $1/20$ of the primary one\footnote{which is $\Delta C / (\gamma_n T_c) \approx 0.74\pm0.02$ where $\gamma_n = C/T$ is evaluated in the normal state~\cite{NishiZakiEA00}}.

In this section we incorporate the above constraints by assuming that $T_c=T_{c1} = 1.48~$K and $T_{c2} = 1.44~$K, and we emphasize that the results below remain fairly insensitive to small variations in $T_{\text{TRSB}}$. The heat capacity is evaluated with 
\begin{equation}
C(T) = 2\sum_{\mu} \sum_{\bo{p}} E_{\mu}(\bo{p}) \f{\D f[E_{\mu} (\bo{p})] }{\D T},
\label{eq:HeatCapacity}
\end{equation}
with quasiparticle energies $E_{\mu}(\bo{p}) = ( \xi_{\mu}(\bo{p})^2 + \lvert \Delta_{\mu}(\bo{p}) \rvert^2 )^{\f{1}{2}}$ and $f(z) = (1+\exp(\beta z))^{-1}$ denoting the Fermi function. Assuming that the order parameter takes the form
\begin{equation}
\begin{aligned}
\Delta(\bo{p}) &= \Delta_{0, a} \left( 1- T/T_{c} \right)^{\f{1}{2}} \Delta_{a}(\bo{p}), \\
&+ i \Delta_{0, b} \left( 1 - T / T_{\text{TRSB}} \right)^{\f{1}{2}} \Delta_{b }(\bo{p})
\end{aligned}
\label{eq:OPansatz}
\end{equation}
(where band indices are suppressed) leads to the following expressions for the ratio of the secondary ($T = T_{\text{TRSB}}$) to the primary ($T = T_{c}$) heat capacity jump~\cite{KivelsonEA20}:
\begin{align}
\eta &= \frac{\Delta C}{T_{\text{TRSB}}}\Big\rvert_{T = T_{\text{TRSB}}} \Big/ \frac{\Delta C}{T_{c}}\Big\rvert_{T = T_{c}} \nonumber \\
&=\left( \f{\Delta_{0, b}}{\Delta_{0, a}} \f{T_{c1}}{T_{\text{TRSB}}} \right)^2 \f{ \big\langle \lvert \Delta_{b}(\bo{p}) \rvert^2 I(\bo{p}) \big\rangle_{\mathrm{FS}} }{ \big\langle \lvert \Delta_{a}(\bo{p}) \rvert^2 \big\rangle_{\mathrm{FS}} }, \label{eq:RatioCJumps} \\
I(\bo{p}) &= \int_{0}^{\infty} \D u~\left( \cosh\left[ \left( u^2 + z(\bo{p})^2 \right)^{\f{1}{2}} \right] \right)^{-2}, \label{eq:Ak} \\
z(\bo{p}) &= \Delta_{0,a} \Delta_{a}(\bo{p})  \left( 1 - T_{\text{TRSB}} / T_{c} \right)^{\f{1}{2}} / (2T_{\text{TRSB}}). \label{eq:zk}
\end{align}
Here the Fermi surface average is evaluated as $\langle f \rangle_{\text{FS}} = \frac{1}{\sum_{\nu}  \rho_{\nu}} \sum_{\mu} \int_{S_{\mu}} \frac{\D \bo{p}}{(2\pi)^2} \frac{f}{v_{\mu}(\bo{p})}$, where $v_{\mu}(\bo{p}) = \lvert \nabla \xi_{\mu}(\bo{p}) \rvert$ is the Fermi velocity, $\rho_{\mu}$ is the density of states (see Eq.~\eqref{eq:DOS}), and where the integral runs over Fermi surface sheet $S_{\mu}$.

\begin{figure}[t!bh]
	\centering
	\subfigure[]{\includegraphics[width=0.90\linewidth]{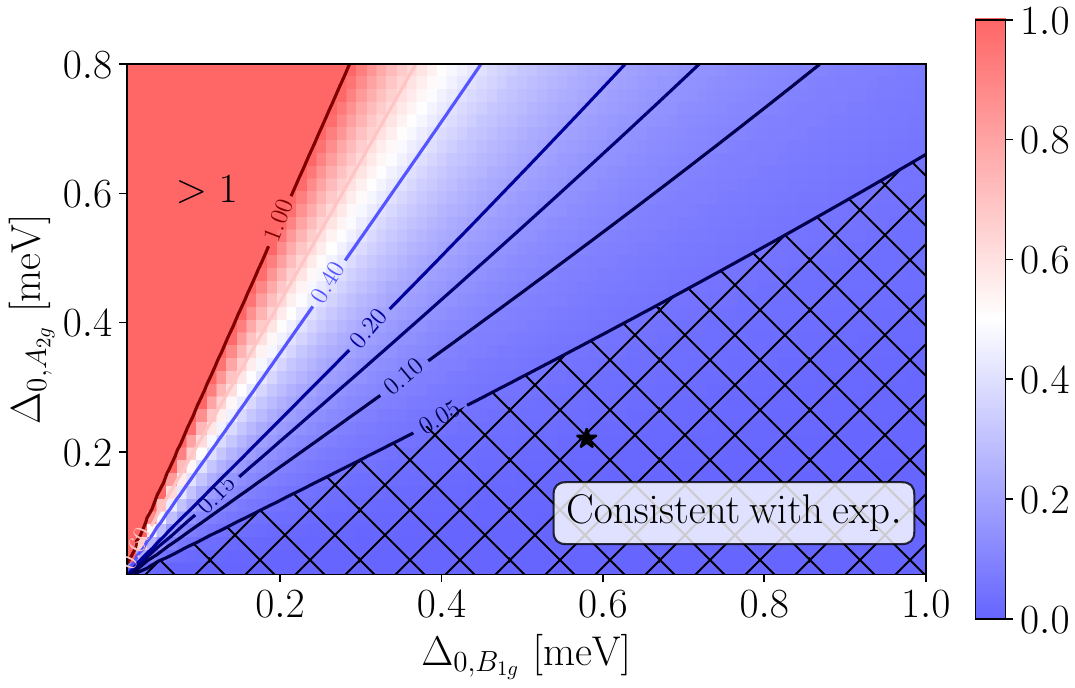} } \quad 
	\subfigure[]{\includegraphics[width=0.90\linewidth]{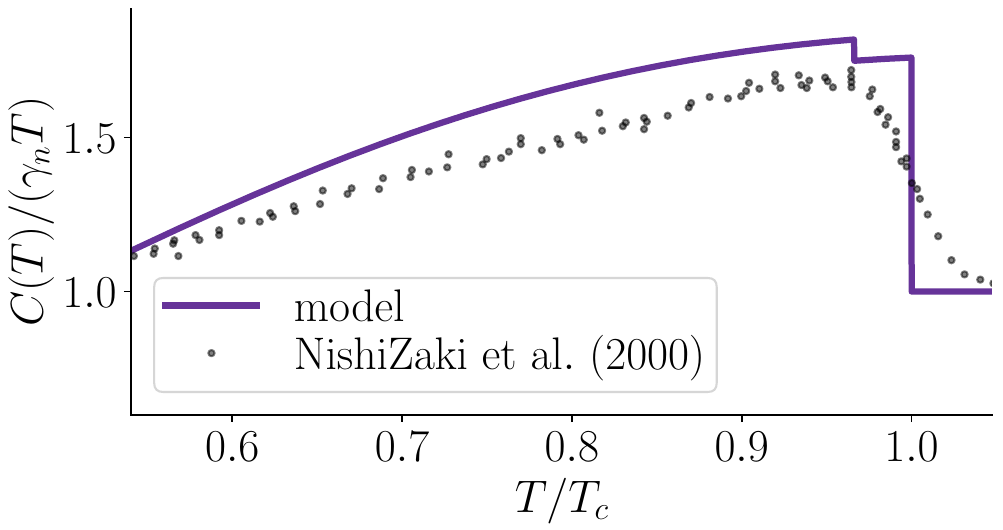} } \quad 
	\subfigure[]{\includegraphics[width=0.90\linewidth]{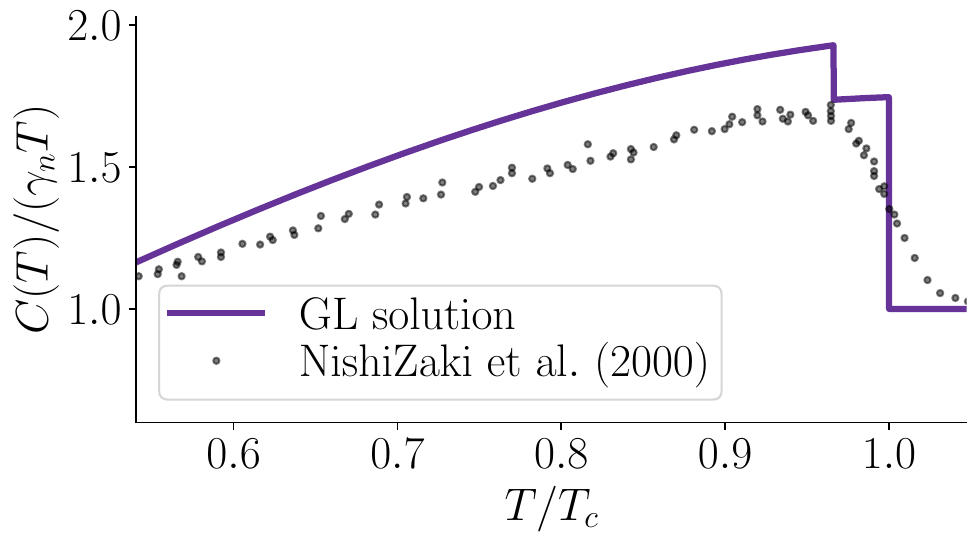} }
\caption{Heat capacity anomaly for a $B_{1g} + i A_{2g}$ order parameter in SRO. Panel (a): Contour lines for the ratio of the second heat capacity jump to the primary heat capacity jump, $\eta$, from Eq.~\eqref{eq:RatioCJumps}, with $T_{\text{TRSB}} = 1.43$~K and $T_c = 1.48$~K using order parameters with three lattice harmonics (see Appendix~\ref{sec:AdvancedOP}). The parameter space consistent with the current experimental threshold, $\eta\lesssim 0.05$, is marked by the cross-hatched region~\cite{GhoshEA20, LiEA19}. For the dispersion we use the (2D) three-band model listed in Appendix~\ref{sec:TBModel}. Panel (b): Specific heat at the point marked with ``$\bigstar$'' in panel (a). Panel (c): Specific heat for the GL solution of Sec.~\ref{sec:MinimizationSRO}. The normalized specific heat per temperature is compared to the data of Ref.~\onlinecite{NishiZakiEA00}.
}
	\label{fig:JumpB1gA2g}
\end{figure}

A colour plot of $\eta$ for the $d+ig$ scenario ($a = B_{1g}$ and $b = A_{2g}$) is shown in Fig.~\ref{fig:JumpB1gA2g} along with the current experimental threshold, $\eta \lesssim 0.05$~\cite{GhoshEA20, LiEA19}. For the order parameters we use those obtained in Ref.~\onlinecite{RoisingEA19}, as well described by the three leading lattice harmonics listed in Appendix~\ref{sec:AdvancedOP}. In Fig.~\ref{fig:JumpB1gA2g} (b) we display the expected specific heat anomaly for parameters close to the experimental threshold, and in Fig.~\ref{fig:JumpB1gA2g} (c) we show the specific heat for the GL solution of Sec.~\ref{sec:MinimizationSRO}. 

The results suggest that the order parameter of Eq.~\eqref{eq:OPansatz} appears consistent with experiments~\cite{GhoshEA20, LiEA19} when $\Delta_{0, A_{2g}} \lesssim 0.6 \Delta_{0, B_{1g}}$. This should be compared with the results of minimizing the GL theory in Fig.~\ref{fig:RGWeights}, for which $\Delta_{0, A_{2g}} \approx 0.71 \Delta_{0, B_{1g}}$ and the second heat capacity jump is greater than the experimental threshold. The result indicates that, in order to be consistent with experiment, a second order parameter component would need to be smaller than that predicted with our theory. Details of the heat capacity calculation are listed in Appendix~\ref{sec:SecondCjump}. 

We note that $T_\textrm{TRSB}$ is not well-known from experiments. However, the size of the jump found in our theory is relatively independent of $T_\textrm{TRSB}$. The importance of the value of $T_\textrm{TRSB}$ is that if it is too close to $T_c$, then the two heat capacity jumps will not be able to be resolved in experiments. We note that when strain is applied, the difference between $T_c$ and $T_\textrm{TRSB}$ increases~\cite{GrinekoEA20}. However, even under applied strain, no second heat capacity jump is observed~\cite{LiEA19}. Our formalism can be extended to the strained case by using the appropriate band structure and band gaps --- we leave this to future work.

An ultrasound spectroscopy experiment recently mapped out the symmetry-resolved elastic tensor of SRO~\cite{GhoshEA20}. The results indicate discontinuous jumps in the compressional elastic moduli ($A_{1g}$) and in only one of the shear elastic moduli ($B_{2g}$). This observation would be consistent with a two-component order parameter where the two components, if belonging to different irreps, form bilinears only in these two channels. As deduced from the direct product table of the irreps in Table~\ref{tab:Representations}, this would be the case for $B_{1g} + i A_{2g}$ and for $A_{1g} + i B_{2g}$. However, only the first of these cases would have symmetry-protected line nodes and thereby offer a robust explanation of the observed heat capacity~\cite{NishiZakiEA00}, heat conductivity~\cite{HassingerEA17}, and STM measurements~\cite{MadhavenEA19}. In Appendix~\ref{sec:AdvancedOP} we examine the order parameter combination $A_{1g} + i B_{2g}$ for completeness. The same conclusion that the heat capacity jump is inconsistent with the experimental data is reached for this order parameter, and the qualitative  features remain fairly insensitive to the precise order parameters used. From a microscopic perspective this latter order parameter, $s'+id$, was recently found to be a viable candidate when including longer-range Coulomb terms in a random phase approximation scheme~\cite{RomerEA21}.

%
%%
%%%
\section{Conclusions}
\label{sec:Conclusion}
%%%
%%
%
In this paper, we have examined the $d+ig$-wave order parameter hypothesis as a candidate model for the superconductivity in SRO. We developed a generic multiband, multi-component Ginzburg--Landau theory for tetragonal lattice systems. We found that the theory favours a $g$ component with magnitude of $71\%$ of the $d$ components at low temperature. On the other hand, the lack of observation of a second heat capacity jump~\cite{LiEA19} requires the $g$-wave component to be less than about 60\% of the $d$-wave component. Together, these two results place tight restrictions on any possible $d+ig$ scenario. Although the $d+ig$ candidate may reconcile a number of experiments, a robust justification for a near-degeneracy of $d$ and $g$-wave order parameters is yet to be found. This outstanding issue is even more apparent when bearing in mind that numerous calculations based on realistic band structures have yet to find a competitive $g$-wave order parameter~\cite{ScaffidiEA14, SteppkeEA17, DodaroEA18, ZhangEA18, GingrasEA19, RomerEA19, RoisingEA19, WangEA19, SuhEA19, ZhiqiangEA20, RomerEA20}.

The continued squeezing of the range of acceptable theoretical scenarios compatible with experiment suggests that further experimental results might need revisiting. In the end, SRO might be more similar to the cuprates than previously thought, and interface experiments have hinted at time-reversal symmetry-invariant superconductivity~\cite{KashiwayaEA19}. One could imagine the scenario of a cuprate-like $d_{x^2-y^2}$-wave order parameter, where the apparent observation of TRSB originates from an anisotropic order parameter component caused by dislocations, magnetic defects, or domain walls~\cite{WillaEA20}, or mechanisms not intrinsically related to superconductivity~\cite{Mazumdar20}.

We also note that yet another order parameter candidate, of the form $d+ip$, has recently been suggested based on the near-degeneracy between even and odd-parity order parameters in the 1D Hubbard model~\cite{Scaffidi20}. This order can potentially reconcile junction experiments suggesting odd-parity order~\cite{NelsonEA04, KidwingiraEA06, AnwarEA17, KashiwayaEA19} with other indications of a nodal $d$-wave~\cite{HassingerEA17, SteppkeEA17, PustogowEA19, MadhavenEA19}. 

The current experimental situation taken at face value appears to leave somewhat exotic options that at least would require further microscopic examination. These new hypotheses warrant careful (re-)examination in hopes of unifying theory and experiment to converge on a solution to the pairing symmetry puzzle in SRO. 

\begin{acknowledgments}
We thank Fabian Jerzembeck, Astrid Tranum R\o{}mer, Catherine Kallin, Thomas Scaffidi, Steven Kivelson, and Alexander Balatsky for useful discussions. We thank Egor Babaev for comments on a previous version of the draft. G.~W.~thanks the Kavli Institute for Theoretical Physics for its hospitality during the graduate fellowship programme. H.~S.~R.~acknowledges support from the Aker Scholarship and VILLUM FONDEN via the Centre of Excellence for Dirac Materials (Grant No.~11744). F.~F.~acknowledges support from the Astor Junior Research Fellowship of New College, Oxford. S.~H.~S.~is supported by EPSRC Grant No.\@ EP/N01930X/1.
\end{acknowledgments}

\bibliographystyle{apsrev4-1}
\bibliography{ReferencesSRO}

\newpage

\begin{appendix}
\onecolumngrid

%
%%
%%%
\section{Specific instances of the GL theory}
\label{sec:MmtmIndepGL}
%%%
%%
%

%
%%
\subsection{General expressions for the Ginzburg--Landau coefficients}
\label{sec:AppNoFluct}
Here we consider the Ginzburg--Landau theory derived in the main text. Assuming $\Delta_{a\mu}(\bo{p})=\Delta_{a}^0\Delta_{a\mu}(\bo{p})$. Here, $\Delta_{a\mu}(\bo{p})$ are the normalized RG gaps and $\Delta_{a}^0$ are the amplitudes of a given irrep, which are the variational parameters over which we want to minimize our free energy. We note that these variational parameters do not depend on the band label $\mu$ since the relative amplitude of the gaps of a given irreps on the different bands is already fixed from our calculation of the RG gaps. In this case the theory of Eqs.~\eqref{eq:GLTheoryOne}, \eqref{eq:SecondOrderF}, \eqref{eq:FourthOrderF} reduces to
\begin{align}
\Delta F&=\Delta F^{(2)}+\Delta F^{(4)}= {\sum_{a}} \tilde\alpha_{a}(T,T_{ca}) \lvert \Delta_{a}^0\rvert^2  + {\sum_{a_i}} \tilde\beta_{\{a_i\}}(T_c) \Delta_{a_1}^{0*} \Delta_{a_2}^{0*} \Delta_{a_3}^{0} \Delta_{a_4}^{0} \Big),
\\
\tilde\alpha_{a}(T,T_{ca}) &= \sum_{n,\mu,\bo{p}} \lvert \Delta_{a\mu}(\bo{p})\rvert^2\bigg(- T   \sum_{n} G_{\mu}(\bo{p}, \omega_n) G_{\mu}(-\bo{p}, -\omega_n)  + T_{ca}  \sum_{n} G_{\mu}(\bo{p}, \omega_n) G_{\mu}(-\bo{p}, -\omega_n) \rvert_{T = T_{ca}}\bigg), \label{eq:AppSecondOrderF} \\
\tilde\beta_{\{a_i\}}(T)&= \f{T}{2} f_{a_1 a_2 a_3 a_4} \Delta_{a_1\mu}(\bo{p})^* \Delta_{a_2\mu}(\bo{p})^* \Delta_{a_3\mu}(\bo{p}) \Delta_{a_4\mu}(\bo{p}) \nonumber \\
&\qquad \times \sum_{n} G_{\mu}(\bo{p}, \omega_n)G_{\mu}(\bo{p}, \omega_n) G_{\mu}(-\bo{p}, -\omega_n)G_{\mu}(-\bo{p}, -\omega_n).\label{eq:AppFourthOrderF}
\end{align}
Here $f_{a_1 a_2 a_3 a_4}$ is as given in Eq.~\eqref{eq:KroneckerCombination}. The frequency sums of Eq.~\eqref{eq:AppSecondOrderF} and \eqref{eq:AppFourthOrderF} are evaluated analytically, and we arrive at the following GL coefficients 
\begin{align}
\Delta F&= {\sum_{a}} \tilde\alpha_{a}(T,T_{ca}) \lvert \Delta_{a}^0\rvert^2  + {\sum_{a_i}} \tilde\beta_{\{a_i\}}(T) \Delta_{a_1}^{0*} \Delta_{a_2}^{0*} \Delta_{a_3}^{0} \Delta_{a_4}^{0} \Big),
\\
\tilde\alpha_{a}(T,T_{ca}) &= - V \sum_{\mu}\int \f{\D \bo{p}}{(2\pi)^d}~\Big( \f{\tanh\left[ \xi_{\mu}(\bo{p}) / (2T) \right] }{2\xi_{\mu}(\bo{p})} -\f{\tanh\left[ \xi_{\mu}(\bo{p}) / (2T_c) \right]}{2\xi_{\mu}(\bo{p})} \Big)\lvert \Delta_{a\mu}(\bo{p})\rvert^2 \\
\tilde\beta_{\{a_i\}}(T) &= f_{a_1a_2a_3a_4} \f{V}{2T^3} \sum_\mu \int \f{\D \bo{p}}{(2\pi)^d}~h( \xi_{\mu}(\bo{p})/ T )\Delta_{a_1\mu}(\bo{p}) \Delta_{a_2\mu}(\bo{p}) \Delta_{a_3\mu}(\bo{p}) \Delta_{a_4\mu}(\bo{p}),
\end{align}
where we have used the fact that the RG gaps are real, and where we introduced
\begin{equation}
h(x) \equiv \frac{\sinh{x}-x}{4x^3(1+\cosh{x})}.
\label{eq:hfunc}
\end{equation}
For two irreps
\begin{align}
\Delta F&= \tilde\alpha_{1}(T,T_{c1}) \lvert \Delta_1\rvert^2+\tilde\alpha_{2}(T,T_{c2}) \lvert \Delta_2\rvert^2 \label{eq:FreeEnergyAppA1} \\&\qquad+ \tilde\beta_{1111}(T) \lvert \Delta_1 \rvert^4+\tilde\beta_{1122}(T) \big( 4\lvert \Delta_1 \rvert^2 \lvert \Delta_2 \rvert^2+\Delta_1^2\Delta_2^{*2}+\Delta_1^{*2}\Delta_2^2 \big) + \tilde\beta_{2222}(T)\lvert \Delta_2 \rvert^4, \nonumber \\
&=\Delta_0^2 \bigg[\tilde\alpha_{1}(T,T_{c1})  +\tilde\alpha_{2}(T,T_{c2})X^2\bigg] + \Delta_0^4\bigg[\tilde\beta_{1111}(T) +2\tilde\beta_{1122}(T) X^2 + \tilde\beta_{2222}(T)X^4\bigg], \nonumber \\
\tilde\alpha_{a}(T,T_{ca}) &= - V \sum_{\mu}\int \f{\D \bo{p}}{(2\pi)^d}~\Big( \f{\tanh\left[ \xi_{\mu}(\bo{p}) / (2T) \right] }{2\xi_{\mu}(\bo{p})} -\f{\tanh\left[ \xi_{\mu}(\bo{p}) / (2T_{ca}) \right]}{2\xi_{\mu}(\bo{p})} \Big)\lvert \Delta_{a\mu}(\bo{p})\rvert^2 \label{eq:AppAlphaNoFluct},  \\
\tilde\beta_{1111}(T) &= \sum_\mu\f{V}{2T^3} \int \f{\D \bo{p}}{(2\pi)^d}~h( \xi_{\mu}(\bo{p})/ T )\Delta_{1\mu}(\bo{p})^4 \label{eq:AppBeta1111}, \\
\tilde\beta_{1122}(T) &= \sum_\mu\f{V}{2T^3} \int \f{\D \bo{p}}{(2\pi)^d}~h( \xi_{\mu}(\bo{p})/ T )\Delta_{1\mu}(\bo{p})^2\Delta_{2\mu}(\bo{p})^2 \label{eq:AppBeta1122}, \\
\tilde\beta_{2222}(T) &= \sum_\mu\f{V}{2T^3} \int \f{\D \bo{p}}{(2\pi)^d}~h( \xi_{\mu}(\bo{p})/ T )\Delta_{2\mu}(\bo{p})^4 \label{eq:AppBeta2222}.
\end{align}
\subsection{Single $s$-wave component}
\label{sec:AppNoFluctSwave}
Here we consider the Ginzburg--Landau theory under the assumptions of (i) a single-component $s$-wave order parameter, and (ii) quadratic bands in 2D. Under these simplifying assumptions we reproduce the results originally obtained by Gor'kov~\cite{Gorkov59}. We assume an $s$-wave order parameter, i.e.~$\Delta_{a\mu}(\bo{p})=\Delta_{A_{1g}\mu} \equiv \Delta_{\mu}$ so the free energy simplifies to 
\begin{align}
\label{eq:AppGLTheoryOne2}\Delta F=\Delta F^{(2)}+\Delta F^{(4)}&= \sum_{\mu} \Big( \tilde\alpha_{\mu}(T,T_c) \lvert \Delta_{\mu}\rvert^2  +  \tilde\beta_{\mu}(T_c) \lvert \Delta_{\mu}\rvert^4 \Big),
\\
\tilde\alpha_{\mu}(T,T_c) &= - T   \sum_{n,\bo{p}} G_{\mu}(\bo{p}, \omega_n) G_{\mu}(-\bo{p}, -\omega_n)  + T_{c}  \sum_{n,\bo{p}} G_{\mu}(\bo{p}, \omega_n) G_{\mu}(-\bo{p}, -\omega_n) \rvert_{T = T_{ca}}, \label{eq:AppSecondOrderF2} \\
\tilde\beta_{\mu}(T)&= \f{T}{2} \sum_{n,\bo{p}} G_{\mu}(\bo{p}, \omega_n)G_{\mu}(\bo{p}, \omega_n) G_{\mu}(-\bo{p}, -\omega_n)G_{\mu}(-\bo{p}, -\omega_n).\label{eq:AppFourthOrderF2}
\end{align}
The frequency sums of Eq.~\eqref{eq:AppSecondOrderF} and \eqref{eq:AppFourthOrderF} are evaluated analytically, and we arrive at the following GL coefficients 
\begin{align}
\tilde\alpha_{\mu}(T,T_c) &= - V \int \f{\D \bo{p}}{(2\pi)^d}~\Big( \f{\tanh\left[ \xi_{\mu}(\bo{p}) / (2T) \right] }{2\xi_{\mu}(\bo{p})} -\f{\tanh\left[ \xi_{\mu}(\bo{p}) / (2T_c) \right]}{2\xi_{\mu}(\bo{p})} \Big) \label{eq:AppAlpha}  \\
\tilde\beta_{\mu}(T) &= \f{V}{2T^3} \int \f{\D \bo{p}}{(2\pi)^d}~h( \xi_{\mu}(\bo{p})/ T ) \label{eq:AppBeta},
\end{align}
upon replacing the momentum sums by $\sum_{\bo{p}} \to V \int \f{\D\bo{p}}{(2\pi)^d}$, where $V$ is the unit cell volume. We have used the fact that the integrands are sharply peaked about the Fermi surface and so we can extend the integral over $\bo{p}$ from an integral over the Fermi surface to an integral over the entire Brillouin zone. Next, we evaluate Eq.~\eqref{eq:AppAlpha} and \eqref{eq:AppBeta} for quadratic bands in 2D, $\xi_{\mu}(\bo{p}) = \f{p^2}{2m_{\mu}}$, with $p = \lvert \bo{p} \rvert$ and $m_{\mu}$ being the effective mass of band $\mu$, and with the Brillouin zone integrals $\int \D \bo{p} \to \int_0^{\infty} \D p~p \int_0^{2\pi} \D\phi$. To evaluate the basic integral of Eq.~\eqref{eq:AppBeta}, $\f{1}{2} \int_0^{\infty} \D u~\f{\sinh{u}-u}{4u^3(1+\cosh{u})}$, we make use of the following series expansions:
\begin{align}
\frac{2}{1+\cosh(x)} &= \cosh^{-2}(x/2) = 4e^{-x} \sum_{n=0}^{\infty} (-1)^n (1+n)e^{-nx}, \label{eq:CoshTaylor} \\
\sinh(x)-x &= \f{\sqrt{\pi}}{2} \sum_{m=1}^{\infty} \f{x^{2m+1}}{4^m \Gamma(m+1) \Gamma(m+\f{3}{2})}, \label{eq:SinhExpansion} \\
\sum_{n=0}^{\infty} (-1)^n \f{1}{(1+n)^x} &= (1-2^{1-x}) \zeta(x). \label{eq:ZetaDef}
\end{align}
By equating the resulting expression for $\beta_{\mu}$ with the result derived by Gor'kov~\cite{Gorkov59}, we find that
\begin{equation}
\begin{aligned}
\f{7\zeta(3)}{32\pi^2} &= \f{\sqrt{\pi}}{8} \sum_{l = 1}^{\infty} (1-2^{3-2l}) \f{\Gamma(2l-1)}{4^l\Gamma(l+1)\Gamma(l+\f{3}{2})} \zeta(2l-2) \Rightarrow \\
\zeta(3) &= \frac{8\pi^2}{7} \sum_{n=0}^{\infty} \frac{\left( 1-2\cdot 2^{-2n} \right) \zeta(2n)}{(2n+1)(2n+2)(2n+3)},
\end{aligned}
\label{eq:Identity}
\end{equation}
where $\zeta(0) = -1/2$. In fact, both of the terms inside the sum of Eq.~\eqref{eq:Identity} individually yield a series expansion for $\zeta(3)$:
\begin{align}
\zeta(3) &= -\frac{8\pi^2}{5} \sum_{n=0}^{\infty} \frac{\zeta(2n)}{(2n+1)(2n+2)(2n+3) 2^{2n} }, \label{eq:NewIdentity2} \\
\zeta(3) &= -\frac{8\pi^2}{3} \sum_{n=0}^{\infty}  \frac{\zeta(2n) }{(2n+1)(2n+2)(2n+3)}. \label{eq:NewIdentity3}
\end{align}
The most rapidly convergent series of the two, Eq.~\eqref{eq:NewIdentity2}, along with plenty of other variations, was discovered by Chen and Srivastava~\cite{ChenSrivastava}. The latter one, however, does not appear to have been discussed in the literature.

Finally, for the coefficients $\tilde{\alpha}_{\mu}$ we assume that $T/T_c-1 \ll 1$ and retain the leading term in a Taylor expansion. The result is
\begin{align}
\tilde\alpha_{\mu} &= \f{\rho_{\mu}}{2}\left(\f{T}{T_c} - 1 \right), \label{eq:AppAlpha2DQuad} \\
\tilde\beta_{\mu} &= \f{\rho_{\mu}}{T_c^2} \f{7\zeta(3)}{32\pi^2}, \label{eq:AppBeta2DQuad}
\end{align}
with $\rho_{\mu} = V m_{\mu}/(2\pi)$ being the density of states and $\zeta$ the Riemann zeta function. This is equivalent to the result of Gor'kov~\cite{Gorkov59}.
Repeating the above exercise for linearly dispersing bands, $\xi_{\mu}(\bo{p}) = v_{\mu} p$, results instead in 
\begin{align}
\tilde\alpha_{\mu} &= \rho_{\mu} \ln 2 \left(\f{T}{T_c} - 1 \right), \label{eq:AppAlpha2DLinear} \\
\tilde\beta_{\mu} &= \f{\rho_{\mu}}{16 T_c^2}, \label{eq:AppBeta2DLinear}
\end{align}
where now $\rho_{\mu} = V T_c/(2\pi v_{\mu}^2)$.

\subsection{Including fluctuations}
\label{sec:AppFluct}
In general the order parameter could depend on the center-of-mass momentum ($|\bo{q}|\ll k_F$), which would allow us to describe spatial fluctuations of the superconducting order parameter. Eq.~\eqref{eq:HSC} would then read:
\begin{equation}
\begin{aligned}
H_{\text{SC}} &= \sum_{\mu, \sigma,a }  {\sum_{\bo{q},\bo{p}}} \big[ \Delta_{a\mu}(\bo{p},\bo{q}) c_{\mu \sigma}^{\dagger}\left(\bo{p}+\bo{q}/2 \right) c_{\mu \bar{\sigma}}^{\dagger} \left(-\bo{p}+\bo{q}/2 \right) + \text{h.c.} \big],
\end{aligned}
\label{eq:HSCApp}
\end{equation}
where $\Delta_{a\mu}(\bo{p},\bo{q})$ is the pseudospin-singlet order parameter of band $\mu$ corresponding to irrep.~$a$. When repeating the steps of Sec.~\ref{sec:Formalism} with the above order parameter we now find the following generalized versions of Eqs.~\eqref{eq:GLTheoryOne}, \eqref{eq:SecondOrderF}, \eqref{eq:FourthOrderF}:
\begin{align}
\label{eq:GLTheoryOneApp}\Delta F=\Delta F^{(2)}+\Delta F^{(4)} &= \sum_{\mu} \Big( {\sum_{a,\bo{p},\bo{q}}} \alpha_{a\mu}(\bo{p},\bo{q},T) \lvert \Delta_{a\mu}(\bo{p},\bo{q})\rvert^2\\&\nonumber\qquad  + {\sum_{a_i,\bo{p}_i,\bo{q}_i}} \beta_{\{a_i\}\mu}(\{\bo{p}_i\},\{\bo{q}_i\},T) \Delta_{a_1\mu}(\bo{p}_1,\bo{q}_1)^* \Delta_{a_2\mu}(\bo{p}_2,\bo{q}_2)^* \Delta_{a_3\mu}(\bo{p}_3,\bo{q}_3) \Delta_{a_4\mu}(\bo{p}_4,\bo{q}_4) \Big),
\\
\alpha_{a\mu}(\bo{p},\bo{q},T) &= - T   \sum_{n} G_{\mu}(\bo{p}+\frac{\bo{q}}{2}, \omega_n) G_{\mu}(-\bo{p}+\frac{\bo{q}}{2}, -\omega_n)  + T_{ca}  \sum_{n} G_{\mu}(\bo{p}, \omega_n) G_{\mu}(-\bo{p}, -\omega_n) \rvert_{T = T_{ca}}, \label{eq:SecondOrderFApp} \\
\beta_{\{a_i\}\mu}(\{\bo{p}_i\},\{\bo{q}_i\},T)&= \f{T}{2} f_{a_1a_2a_3a_4} \delta_{\bo{p}_1+\frac{\bo{q}_1}{2},\bo{p}_3+\frac{\bo{q}_3}{2}}\delta_{\bo{p}_1-\frac{\bo{q}_1}{2},\bo{p}_4-\frac{\bo{q}_4}{2}}\delta_{\bo{p}_2-\frac{\bo{q}_2}{2},\bo{p}_3-\frac{\bo{q}_3}{2}}\delta_{\bo{p}_2+\frac{\bo{q}_2}{2},\bo{p}_4+\frac{\bo{q}_4}{2}}\label{eq:FourthOrderFApp}
\\ &\quad \times \sum_{n} G_{\mu}(\bo{p}_1+\frac{\bo{q}_1}{2}, \omega_n)G_{\mu}(\bo{p}_2+\frac{\bo{q}_2}{2}, \omega_n) G_{\mu}(-\bo{p}_3+\frac{\bo{q}_3}{2}, -\omega_n)G_{\mu}(-\bo{p}_4+\frac{\bo{q}_4}{2}, -\omega_n), \nonumber \\
f_{a_1a_2a_3a_4} &\equiv \delta_{a_1a_3}\delta_{a_2a_4}+\delta_{a_1a_4}\delta_{a_2a_3}+\delta_{a_1a_2}\delta_{a_3a_4}-2\delta_{a_1a_2}\delta_{a_2a_3}\delta_{a_3a_4}. \label{eq:KroneckerCombinationApp}
\end{align}

\section{Tight-binding model}
\label{sec:TBModel}
%%%
%%
%
We consider an effective yet accurate two-dimensional, three-band, tight-binding model for Sr$_2$RuO$_4$,
\begin{equation}
H_{\text{K}} = \sum_{\bo{k}, s} \bo{\psi}^{\dagger}_s(\bo{k}) h_s(\bo{k}) \bo{\psi}_s(\bo{k}),
\label{eq:TBHamiltonian}
\end{equation}
where $\bo{\psi}_s(\bo{k}) = [c_{ xz, s}(\bo{k}), \hspace{1mm} c_{yz, s}(\bo{k}), \hspace{1mm} c_{xy, -s}(\bo{k})]^T$ and where  $s \in \lbrace \uparrow, \downarrow \rbrace$ denotes spin and $a\in\{xz,yz,xy\}$ denotes the $d$-orbitals of the Ruthenium atoms in SRO which are relevant close to the Fermi energy. The matrix $h_s(\bo{k})$ is well approximated by the $3\times 3$ block diagonal matrix
\begin{equation}
h_s(\bo{k}) = \begin{pmatrix}
\varepsilon_{xz}(\bo{k}) & - i s\lambda &  i\lambda \\
 i s\lambda & \varepsilon_{yz}(\bo{k}) &  - s\lambda \\ - i\lambda & - s\lambda & \varepsilon_{xy}(\bo{k})
\end{pmatrix},
\label{eq:KineticHamiltonian}
\end{equation}
where spin-orbit coupling is parametrized by $\lambda$, and the above energies are given by
\begin{align}
\varepsilon_{\mathrm{1D}}(k_{\parallel}, k_{\perp}) &= - 2t_1 \cos{k_{\parallel}} - 2t_2 \cos(2 k_{\parallel}) - 2t_3 \cos{k_{\perp}} - 4t_4 \cos{k_{\parallel}} \cos{k_{\perp}}  \nonumber \\
& \hspace{15pt}  - 4t_5 \cos(2 k_{\parallel}) \cos{k_{\perp}} - 2t_6 \cos(3k_{\parallel})  - \mu_{\mathrm{1D}}, \label{eq:1DHopping} \\
\varepsilon_{\mathrm{2D}}(k_x, k_y) &= - 2\bar{t}_1 \left[ \cos{k_x} + \cos{k_y} \right] - 4\bar{t}_2 \cos{k_x} \cos{k_y}  \nonumber \\
& \hspace{15pt} - 4\bar{t}_3 \left[ \cos(2 k_x) \cos(k_y) + \cos(2 k_y) \cos(k_x) \right] - 4\bar{t}_4 \cos(2k_x) \cos(2k_y)   \nonumber \\
& \hspace{15pt} - 2\bar{t}_5 \left[ \cos(2 k_x) + \cos(2 k_y) \right]  - 4\bar{t}_6 \left[ \cos(3 k_x) \cos(k_y) + \cos(3 k_y) \cos(k_x) \right]  \nonumber \\
&\hspace{15pt} - 4\bar{t}_7 \left[ \cos(3 k_x) + \cos(3 k_y) \right] - \mu_{\mathrm{2D}}, \label{eq:2DHopping} 
\end{align}
with the identifications $\varepsilon_{xz}(\bo{k}) = \varepsilon_{\mathrm{1D}}(k_x, k_y)$, $\varepsilon_{yz}(\bo{k}) = \varepsilon_{\mathrm{1D}}(k_y, k_x)$, and $\varepsilon_{xy}(\bo{k}) = \varepsilon_{\mathrm{2D}}(k_x, k_y)$. We extract the tight-binding parameters, via the Wannier functions for the Ru $t_{2g}$ electron orbitals, resulting from a fully relativistic density functional theory calculation which includes spin-orbit coupling~\cite{SteppkeEA17, Rosner}. Extracted parameters are listed in Table~\ref{tab:HoppingParameters1} and \ref{tab:HoppingParameters2}.
\begin{table}
\centering
\caption{Tight-binding parameters for Eqs.~\eqref{eq:KineticHamiltonian} and \eqref{eq:1DHopping}.}
\begin{tabular}{p{2.5cm} p{1.3cm} p{1.3cm} p{1.3cm} p{1.3cm} p{1.3cm} p{1.3cm} p{1.3cm} p{1.3cm}}  
\toprule
Parameter & $t_1$ & $t_2$ & $t_3$ & $t_4$ & $t_5$ & $t_6$ & $\mu_{\mathrm{1D}}$ & $\lambda$  \\ \hline
Value [meV] & $296.2$ & $-57.3$ & $52.6$ & $-15.6$ & $-15.1$ & $-11.6$ & $315.6$ & $-50.7$  \\ \hline \hline
\label{tab:HoppingParameters1}
\end{tabular}
\end{table}
\begin{table}
\centering
\caption{Tight-binding parameters for Eq.~\eqref{eq:2DHopping}.}
\begin{tabular}{p{2.5cm} p{1.4cm} p{1.4cm} p{1.4cm} p{1.4cm} p{1.4cm} p{1.4cm} p{1.4cm} p{1.4cm}}  
\toprule
Parameter & $\bar{t}_1$ & $\bar{t}_2$ & $\bar{t}_3$ & $\bar{t}_4$ & $\bar{t}_5$ & $\bar{t}_6$ & $\bar{t}_7$ & $\mu_{\mathrm{2D}}$  \\ \hline
Value [meV] & $369.5$ & $123.2$ & $20.4$ & $13.9$ & $-6.0$ & $3.2$ & $2.8$ & $432.5$ \\ \hline \hline
\label{tab:HoppingParameters2}
\end{tabular}
\end{table}
\begin{figure}[t!bh]
	\centering
	\subfigure[]{\includegraphics[width=0.35\linewidth]{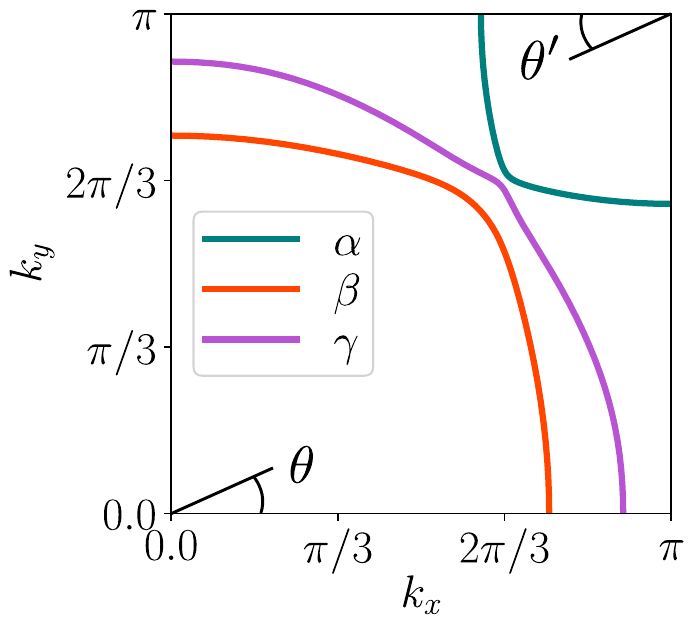} } \quad \subfigure[]{\includegraphics[width=0.40\linewidth]{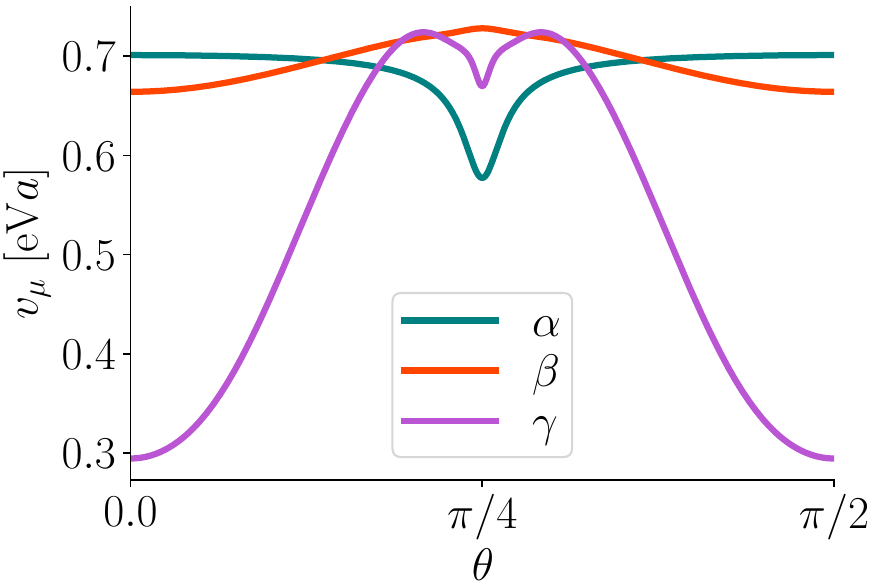} }
\caption{(a) Fermi surface sheets resulting from the model of Eq.~\eqref{eq:TBHamiltonian}, and (b) the Fermi velocity as a function of the in-plane angle $\theta$ ($\theta'$) for bands $\beta$ and $\gamma$ ($\alpha$), \emph{cf}.~Ref.~\onlinecite{TamaiEA19}.}
	\label{fig:FSVelocity}
\end{figure}
We now diagonalize the single-particle Hamiltonian by going from the orbital/spin basis with electron operators $c_{as}(\bo{k})$ to the band/pseudospin basis with electron operators $c_{\mu\sigma}(\bo{k})$, where $\mu\in\{\alpha,\beta,\gamma\}$ denotes the three bands of SRO which intersect the Fermi energy and $\sigma\in\{\Uparrow,\Downarrow\}$ denotes pseudospin. In the band/pseudospin basis the tight-binding Hamiltonian is diagonal:
\begin{equation}
    H_K= \sum_{\mu, \sigma, \bo{k}} \xi_{\mu}(\bo{k}) c_{\mu\sigma}^{\dagger}(\bo{k}) c_{\mu \sigma}(\bo{k}).
\end{equation}
The resulting Fermi surface sheets and Fermi velocities are shown in Fig.~\ref{fig:FSVelocity}. A recent high-resolution ARPES experiment~\cite{TamaiEA19} deduced the Fermi velocities at the Fermi level for bands $\beta$ and $\gamma$. Compared to this experiment the effective model used here is seen to capture the correct behaviour for $v_{\gamma}$, but the behaviour of $v_{\beta}$ (the curvature) is slightly off. Quantitatively, however, this discrepancy is too small to affect the results obtained here in any noticeable way. This was checked explicitly by comparing the results for $\eta$ in Eq.~\eqref{eq:RatioCJumps} to those obtained with $v_{\mu}(\bo{k}) = 1$~eV$a$ fixed.

Serving as a supplementary calculation the relative band densities at the Fermi level produced with this model are $\rho_{\mu}/\rho_{\text{tot}} = 0.163,~0.314,~0.523$ for $\mu = \alpha, \beta, \gamma$, respectively, with
\begin{equation}
\rho_{\mu} = \int_{S_{\mu}} \f{\D \bo{k}}{(2\pi)^2} \frac{1}{\lvert \nabla \xi_{\mu}(\bo{k}) \rvert},
\label{eq:DOS}
\end{equation}
where $S_{\mu}$ is the Fermi surface sheet corresponding to band $\mu$. These values may be compared to those obtained with other models.
\section{Comparison of Hamiltonians}
\begin{figure}[t!hb]
    \centering
    \includegraphics[width=0.5\textwidth]{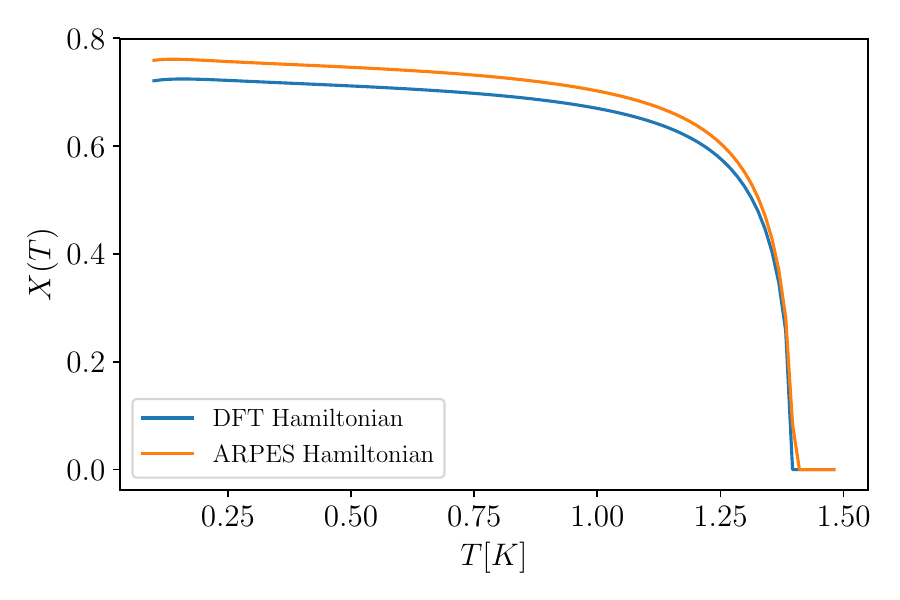}
    \caption{Comparison of the numerical result for $X(T)$ with the order parameter combination $\Delta(\bo{p}) = \Delta_0(T) [\Delta_{B_{1g} \mu}(\bo{p}) + i X(T) \Delta_{A_{2g} \mu}(\bo{p}) ]$ using two different Hamiltonians. The DFT Hamiltonian is described in App.~\ref{sec:TBModel}, the ARPES Hamiltonian is described in Ref.~\onlinecite{RoisingEA19}. The results are very similar.}
    \label{fig:comparing_Hams}
\end{figure}

Fig.~\ref{fig:comparing_Hams} shows the comparison of the results for $X(T)$ using two different Hamiltonians for the bandstructure of SRO.

%
%%
%%%
\section{Order parameters and further plots}
\label{sec:AdvancedOP}
\begin{figure}[t!bh]
	\centering
	\subfigure[]{\includegraphics[width=0.40\linewidth]{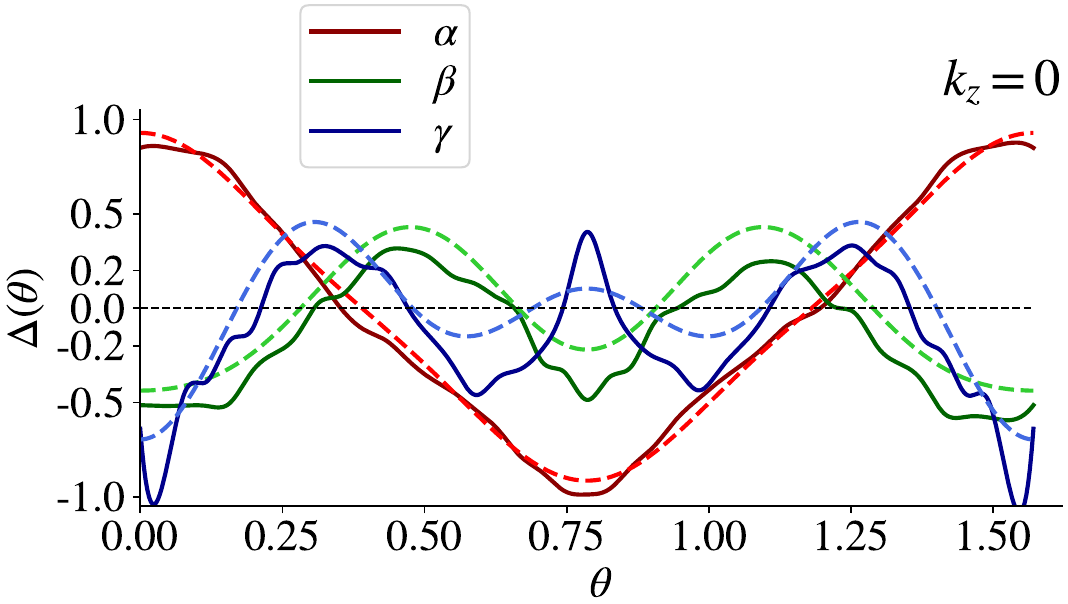} } \quad \subfigure[]{\includegraphics[width=0.40\linewidth]{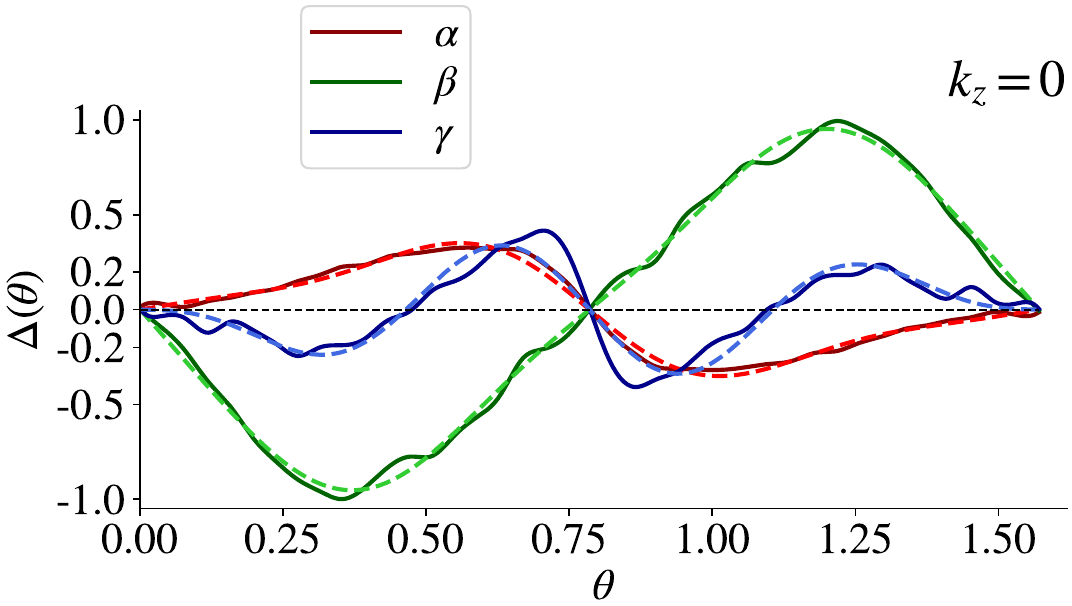} } \quad
	\subfigure[]{\includegraphics[width=0.40\linewidth]{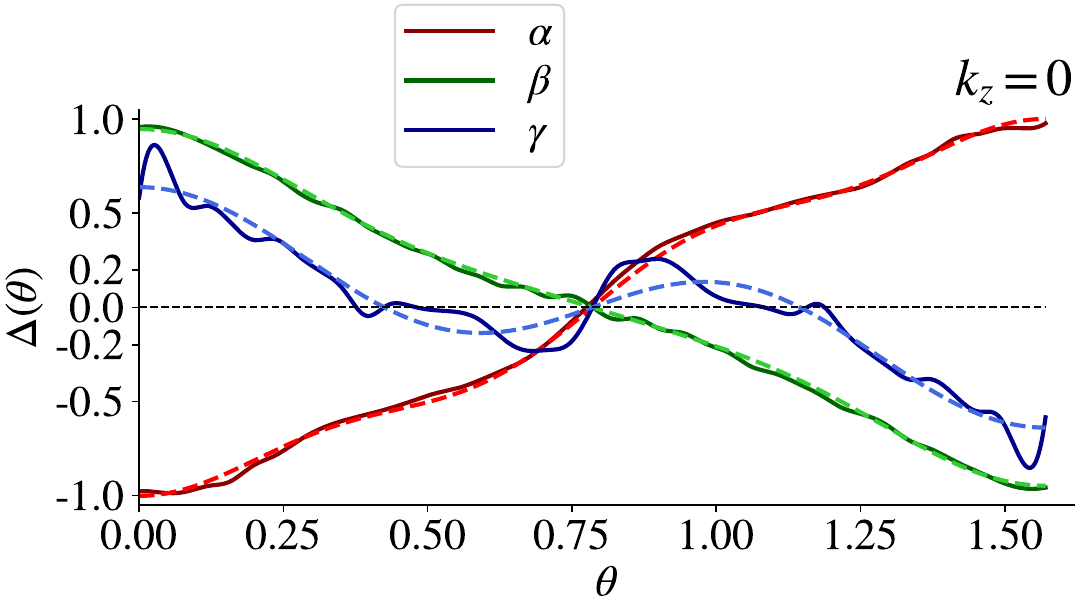} } \quad \subfigure[]{\includegraphics[width=0.40\linewidth]{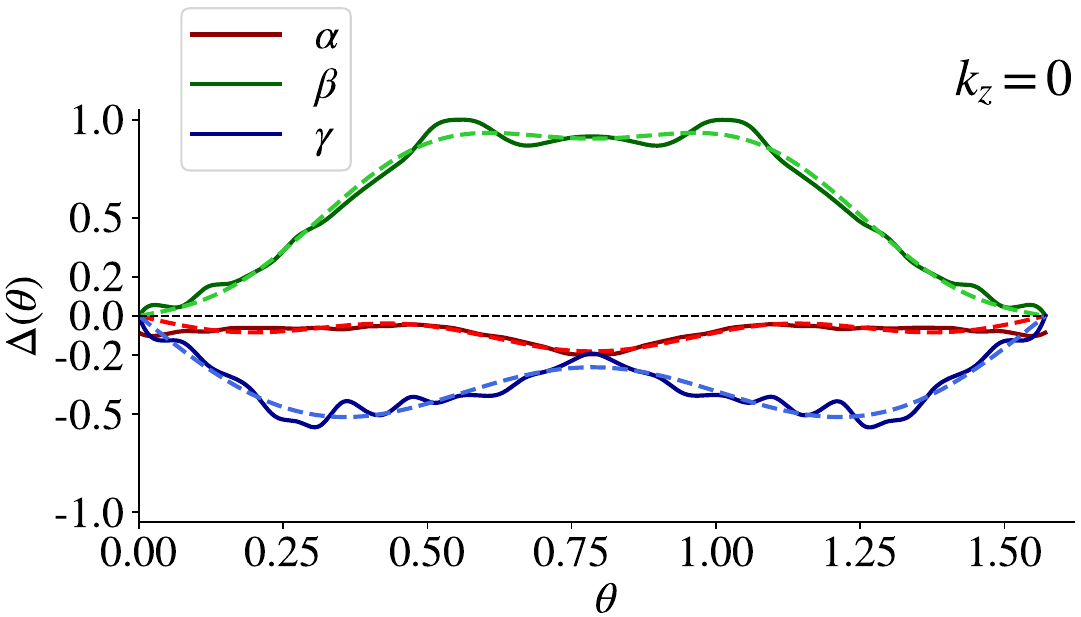} }
\caption{Order parameters from Ref.~\onlinecite{RoisingEA19} for $J/U = 0.20$ (full lines) and lattice harmonics fits (dashed lines) for irreps.~(a) $A_{1g}$, (b) $A_{2g}$, (c) $B_{1g}$, and (d) $B_{2g}$.}
	\label{fig:OPs}
\end{figure}

%%%
%%
%
The lattice harmonics for the order parameters of Tab.~\ref{tab:Representations} are in general band-dependent. The microscopically obtained gap structures of Ref.~\onlinecite{RoisingEA19} (at $J/U = 0.20$) can be well described by the lowest three lattice harmonics. The result of a fitting procedure of the order parameters of symmetries $A_{1g}$, $A_{2g}$, $B_{1g}$, and $B_{2g}$ are listed in Table~\ref{tab:A1gA2gfit} and \ref{tab:B1gB2gfit}, and shown in Fig.~\ref{fig:OPs}. We note that these order parameters strictly were obtained for a different band structure (i.e.~a three-dimensional dispersion based on a band structure fit) than that described in App.~\ref{sec:TBModel}, though the quantitative differences are small in terms of the Fermi surface physics. For the purpose of quantifying the heat capacity anomaly in a realistic model we take these order parameters as reasonable input for the Ginzburg--Landau minimization procedure, while noting that the framework developed here is general and may be employed for other input order parameters in future work.
\begin{table}
\centering
\caption{Lattice harmonics coefficients of the $A_{1g}$ and $A_{2g}$ order parameter (see Tab.~\ref{tab:Representations}) obtained at $J/U = 0.20$ of Ref.~\onlinecite{RoisingEA19}, normalized such that $\max_{\theta, \mu} \Delta_{a \mu}(\theta) = 1$.}
\begin{tabular}{p{1.0cm} p{1.3cm} p{1.3cm} p{1.3cm}}  
\toprule
$\mu$ & $a_{1,\mu}$ & $a_{2,\mu}$ & $a_{3,\mu}$ \\ \hline
$\alpha$ & $+0.855$ & $+0.007$ & $+0.067$ \\ \hline
$\beta$ & $-0.225$ & $-0.329$ & $+0.116$ \\ \hline
$\gamma$ & $-0.097$ & $-0.296$ & $-0.303$ \\ \hline \hline
\end{tabular}
\quad
\begin{tabular}{p{1.0cm} p{1.3cm} p{1.3cm} p{1.3cm}}  
\toprule
$\mu$ & $b_{0,\mu}$ & $b_{1,\mu}$ & $b_{2,\mu}$ \\ \hline
$\alpha$ & $+0.269$ & $-0.127$ & $+0.038$ \\ \hline
$\beta$ & $-0.895$ & $-0.062$ & $+0.052$ \\ \hline
$\gamma$ & $-0.022$ & $-0.219$ & $+0.150$ \\ \hline \hline
\end{tabular}
\label{tab:A1gA2gfit}
\end{table}
\begin{table}
\centering
\caption{Same as in Table~\ref{tab:A1gA2gfit} but for symmetry channels $B_{1g}$ and $B_{2g}$ (see Tab.~\ref{tab:Representations}).}
\begin{tabular}{p{1.0cm} p{1.3cm} p{1.3cm} p{1.3cm}}  
\toprule
$\mu$ & $c_{0,\mu}$ & $c_{1,\mu}$ & $c_{2,\mu}$ \\ \hline
$\alpha$ & $-0.912$ & $-0.011$ & $-0.078$ \\ \hline
$\beta$ & $+0.783$ & $+0.143$ & $+0.022$ \\ \hline
$\gamma$ & $+0.358$ & $+0.288$ & $-0.007$ \\ \hline \hline
\end{tabular}
\quad
\begin{tabular}{p{1.0cm} p{1.3cm} p{1.3cm} p{1.3cm}}  
\toprule
$\mu$ & $d_{0,\mu}$ & $d_{1,\mu}$ & $d_{2,\mu}$ \\ \hline
$\alpha$ & $-0.120$ & $+0.010$ & $-0.051$ \\ \hline
$\beta$ & $+0.943$ & $-0.060$ & $-0.099$ \\ \hline
$\gamma$ & $-0.492$ & $-0.230$ & $+0.0004$ \\ \hline \hline
\end{tabular}
 \label{tab:B1gB2gfit}
\end{table}

\begin{figure}[t!bh]
	\centering
	\subfigure[]{\includegraphics[width=0.42\linewidth]{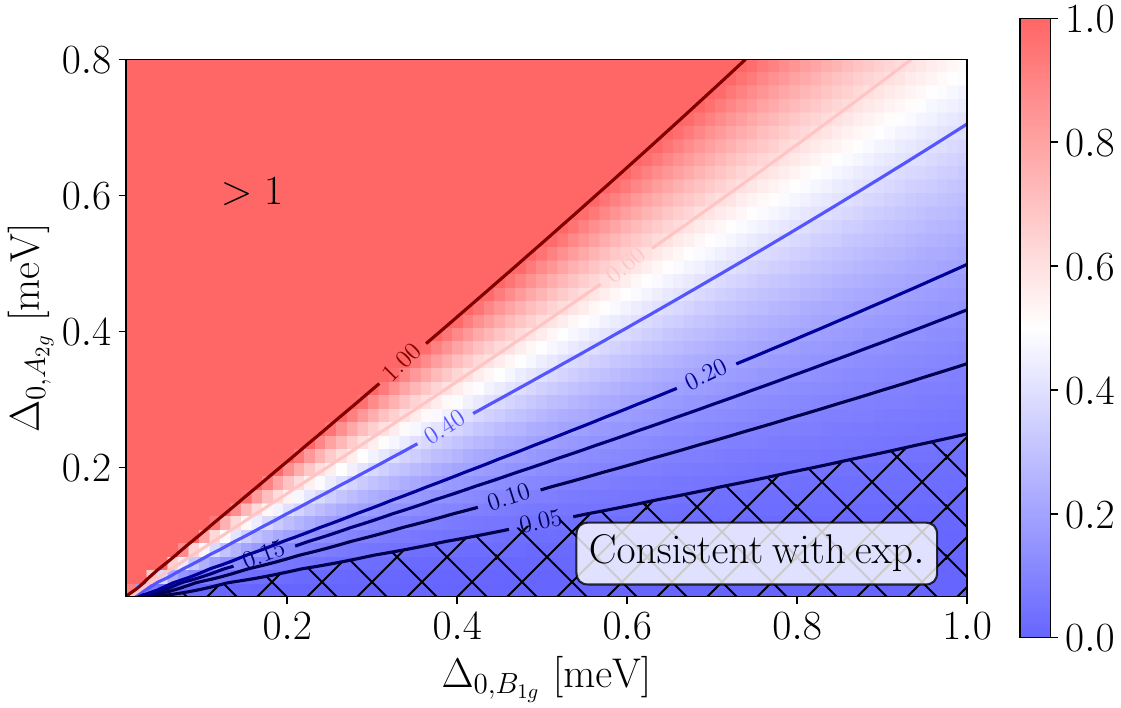} } \quad \subfigure[]{\includegraphics[width=0.42\linewidth]{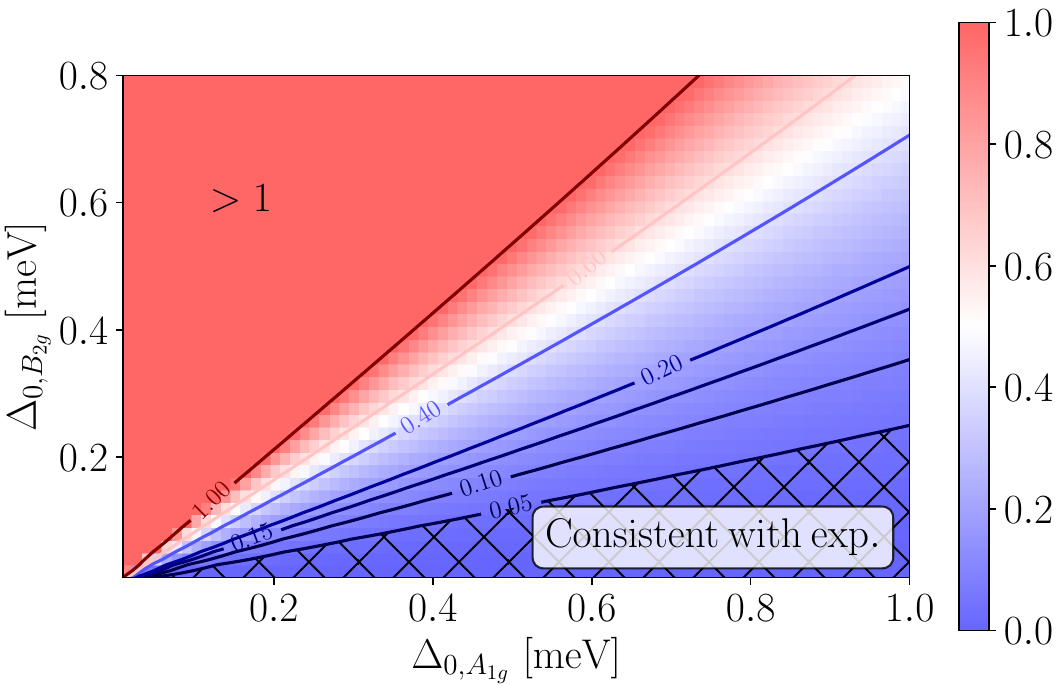} }
\caption{The same as described in the caption of  Fig.~\ref{fig:JumpB1gA2g} but using only the leading lattice harmonics from Table~\ref{tab:Representations} for (a) a $B_{1g} + i A_{2g}$ order parameter, and (b) a $A_{1g} + i B_{2g}$ order parameter.}
	\label{fig:JumpB1gA2gSimple}
\end{figure}

\begin{figure}[t!bh]
	\centering
	\includegraphics[width=0.45\linewidth]{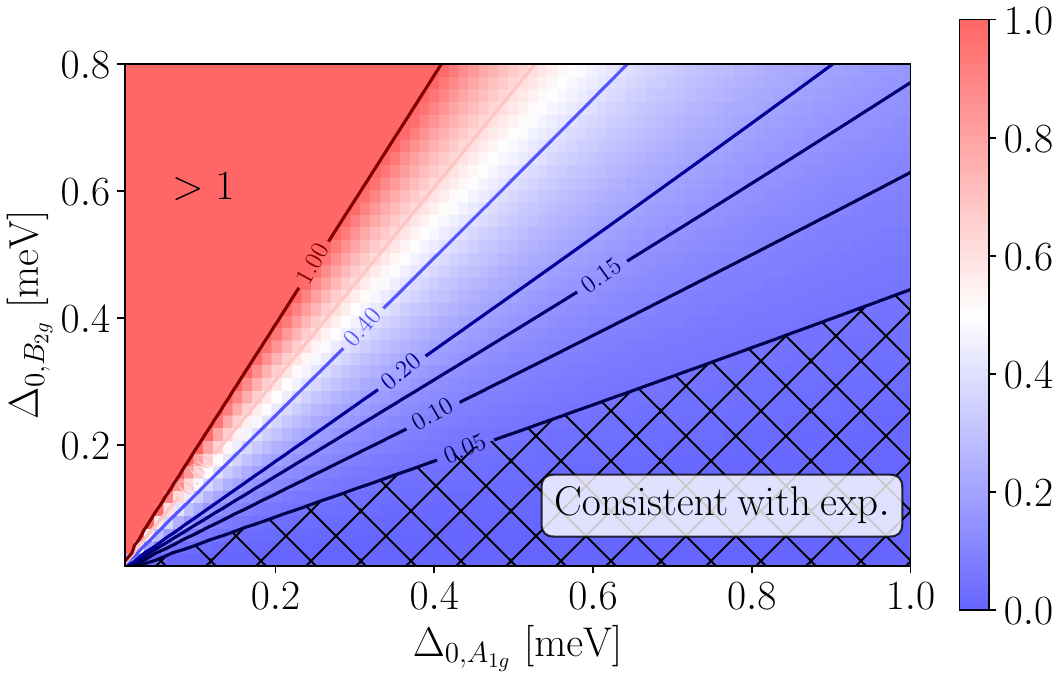}
\caption{The same as described in the caption of  Fig.~\ref{fig:JumpB1gA2g} but here for an order parameter of the form $A_{1g} + i B_{2g}$, using the advanced order parameters with the three leading lattice harmonics from Table~\ref{tab:A1gA2gfit} and \ref{tab:B1gB2gfit}.}
	\label{fig:JumpAdditionalOPS}
\end{figure}

\begin{figure}[t!bh]
	\centering
	\includegraphics[width=0.5\linewidth]{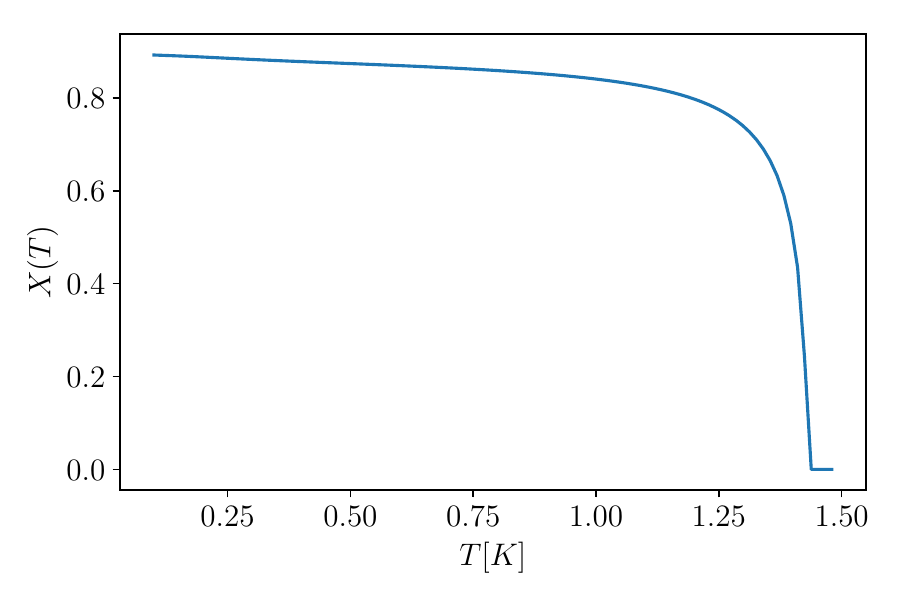}
\caption{The weight $X(T)$, with an order parameter of the form $\Delta(\bo{p}) = \Delta_0(T) [\Delta_{A_{1g} \mu}(\bo{p}) + i X(T) \Delta_{B_{2g} \mu}(\bo{p}) ]$ for $T_{c1}=1.48$~K and $T_{c2} = 1.44$~K. This result is obtained using the GL coefficients Eqs.~\eqref{eq:AppAlpha} and \eqref{eq:AppBeta}.}
	\label{fig:RGWeights2}
\end{figure}

To supplement the results for the heat capacity ratio $\eta$ shown in Fig.~\ref{fig:JumpB1gA2g}, Fig.~\ref{fig:JumpB1gA2gSimple} shows the result of the same calculation using only the leading lattice harmonic. Comparing the two figures shows that including more structure in the order parameter increases the size of the parameter space compatible with experiment~\cite{LiEA19}.
Moreover, Fig.~\ref{fig:JumpAdditionalOPS} shows the outcome of the same calculation for the alternative order parameter combination $A_{1g} + i B_{2g}$, using the three leading lattice harmonics from Table~\ref{tab:A1gA2gfit} and \ref{tab:B1gB2gfit} here, respectively. For this order parameter combination the results indicate compatibility with experiments when $\Delta_{0,B_{2g}} \lesssim 0.4 \Delta_{0, A_{1g}}$.
Finally, Fig.~\ref{fig:RGWeights2} shows the results of minimizing the GL theory of Sec.~\ref{sec:Derivation} for the $A_{1g} + i B_{2g}$ order parameter.
%

%
%%
%%%
\section{Second heat capacity jump}
\label{sec:SecondCjump}
%%%
%%
%
The heat capacity jump at $T_\textrm{TRSB}$ is determined by the discontinuity in $\partial|\Delta|^2/\partial T$, as seen from the normalized expression (the constant $\gamma_n$ below is defined such that $1 = C(T)/(T \gamma_n)\rvert_{T > T_c}$)~\cite{Sigrist05}
\begin{equation}
\frac{C(T)}{T \gamma_n} = \frac{3}{4\pi^2(k_B T)^3 } \int_{- \infty}^{\infty} \D \xi \hspace{1mm} \Big\langle \f{\xi^2 + \lvert \Delta_{\mu}(\bo{p}, T) \rvert^2 - \f{T}{2} \f{\partial \lvert \Delta_{\mu}(\bo{p}, T) \rvert^2}{\partial T} }{\cosh^2(\f{E_{\mu}(\bo{p})}{2k_B T})} \Big\rangle_{\mathrm{FS}},
\label{eq:SpecificHeatNormalized}
\end{equation}
where the Fermi surface average is evaluated as 
\begin{equation}
\langle A \rangle_{\text{FS}} = \frac{1}{\sum_{\nu}  \rho_{\nu}} \sum_{\mu} \int_{S_{\mu}} \frac{\D \bo{p}}{(2\pi)^2} \frac{A}{v_{\mu}(\bo{p})},
\label{eq:NewAverage}
\end{equation}
where $v_{\mu}(\bo{p}) = \lvert \nabla \xi_{\mu}(\bo{p}) \rvert$ is Fermi velocity of band $\mu$. Assuming a gap function of the following form 
\begin{equation}
    \Delta_{\mu}(\bo{p},T)=\Delta_0(T) [\Delta_{1\mu}(\bo{p})+iX(T)\Delta_{2\mu}(\bo{p})],
\end{equation}
the free energy of Eq.~\eqref{eq:FreeEnergyAppA1} is minimized by
\begin{equation}
    X(T)^2=  \begin{cases}
        \frac{\tilde\alpha_1(T,T_{c1})\tilde\beta_{1122}(T)-\tilde\alpha_2(T,T_{c2})\tilde\beta_{1111}(T)}{\tilde\alpha_2(T,T_{c2})\tilde\beta_{1122}(T)-\tilde\beta_{2222}(T)\tilde\alpha_1(T,T_{c1})} & \text{for } T<T_\textrm{TRSB}\\
        0 & \text{for } T>T_\textrm{TRSB},
        \end{cases}
\end{equation}
and
\begin{equation}
    \Delta_0(T)^2 =  \begin{cases}
        -\frac{1}{2}\frac{\tilde\alpha_2(T,T_{c2})\tilde\beta_{1122}(T)-\tilde\beta_{2222}(T)\tilde\alpha_1(T,T_{c1})}{\tilde\beta_{1122}(T)^2-\tilde\beta_{2222}(T)\tilde\beta_{1111}(T)} & \text{for } T<T_\textrm{TRSB}\\
        -\frac{\tilde\alpha_1(T,T_{c1})}{2\tilde\beta_{1111}(T)} & \text{for } T_\textrm{TRSB}<T<T_c\\
        0 & \text{for } T>T_c,
        \end{cases}
\end{equation}
one can derive
\begin{equation}
    \frac{\partial|\Delta_{\mu}(\bo{p},T)|^2}{\partial T}\bigg|^{T_\textrm{TRSB}+\varepsilon}_{T_\textrm{TRSB}-\varepsilon}=\frac{1}{2}\frac{\frac{\partial\tilde\alpha_1(T,T_{c1})}{\partial T}\tilde\beta_{1122}(T)-\frac{\partial\tilde\alpha_2(T,T_{c2})}{\partial T}\tilde\beta_{1111}(T)}{\tilde\beta_{1122}(T)^2-\tilde\beta_{2222}(T)\tilde\beta_{1111}(T)}\bigg(-\frac{\tilde\beta_{1122}(T)}{\tilde\beta_{1111}(T)}\Delta_{1\mu}(\bo{p})^2+\Delta_{2\mu}(\bo{p})^2\bigg).
    \label{eq:disc}
\end{equation}

%
%%
%%%

\end{appendix}

\end{document}